\documentclass[aps,11pt,twocolumn,letterpaper,superscriptaddress,amsmath,amsfonts]{revtex4-2}
\usepackage{amssymb}
\usepackage{amsbsy}
\usepackage{amsmath}
\usepackage{graphicx}
\usepackage{graphics}
\usepackage{setspace}
\usepackage{array}
\usepackage{color}
\usepackage{pdfpages}
\usepackage{fontenc}
\usepackage{textcomp}
\usepackage{float}
\usepackage{bm}
\usepackage{cancel}
\usepackage{import}

\usepackage{pdfpages}
\usepackage{etoolbox} 

\makeatletter
\patchcmd{\@outputpage@head}{\@ifx{\LS@rot\@undefined}{}{\LS@rot}}{}{}{}
\makeatother

\usepackage[body={17.cm,22.5cm},top={3cm},left={2.5cm}]{geometry}

\let\a=\alpha \let\b=\beta \let\g=\gamma \let\d=\partial 
   \let\k=\kappa
\let\l=\lambda     
\let\s=\sigma   

\let\w=\omega \let\G=\Gamma \let\D=\Delta  
   
 \let\W=\Omega

\def\nn{\nonumber}

\def\bpm{\begin{pmatrix}}
\def\epm{\end{pmatrix}}
\def\be{\begin{equation}}
\def\ee{\end{equation}}
\def\bea{\begin{eqnarray}}
\def\eea{\end{eqnarray}}
\def\ba{\begin{array}}
\def\ea{\end{array}}
\def\del{\delta}

\def\td{\tilde}
\def\wtd{\widetilde}

\def\etal{\emph{et al.}}
\def\ep{{\epsilon}}
\def\vep{{\varepsilon}}

\def\bfr{\mathbf{r}}

\newcommand{\bk}{{\bf k}}

\newcommand{\bp}{{\bf p}}
\newcommand{\bq}{{\bf q}}

\newcommand{\bQ}{{\bf Q}}

\newcommand{\mathsym}[1]{{}}
\newcommand{\unicode}[1]{{}}

\newcommand{\besb}{\begin{subequations}}
\newcommand{\eesb}{\end{subequations}}
\newcommand{\beal}{\begin{align}}
\newcommand{\eeal}{\end{align}}

\begin{document}
\title{Theory of topological exciton insulators and condensates in flat Chern bands}
\author{Hong-Yi~Xie} \email{hongyi.xie-1@ou.edu}
\affiliation{Department of Physics and Astronomy, Center for Quantum Research and Technology, University of Oklahoma, Norman, Oklahoma 73069, USA}
\author{Pouyan Ghaemi}
\affiliation{Department of Physics, City University of New York, New York, 10031, USA}
\affiliation{Physics Program, Graduate Center of City University of New York, NY 10031, USA}
\author{Matteo Mitrano}
\affiliation{Department of Physics, Harvard University,  Cambridge, Massachussetts, 02138, USA}
\author{Bruno Uchoa} \email{uchoa@ou.edu}
\affiliation{Department of Physics and Astronomy, Center for Quantum Research and Technology, University of Oklahoma, Norman, Oklahoma 73069, USA}
\date \today

\begin{abstract}
Excitons are the neutral quasiparticles that form when Coulomb interactions create bound states between electrons and holes.  
Due to their bosonic nature, excitons are expected to condense and exhibit superfluidity at sufficiently low temperatures. 
In interacting Chern insulators, excitons may inherit the nontrivial topology and quantum geometry from the underlying electron wavefunctions.
We theoretically investigate the excitonic bound states and superfluidity in flat-band insulators pumped with light. 
We find that the exciton wavefunctions exhibit vortex structures in momentum space, with the total vorticity being equal to the difference of Chern numbers between the conduction and valence bands. 
Moreover, both the exciton binding energy and the exciton superfluid density are proportional to the Brillouin-zone average of the quantum metric and the Coulomb potential energy per unit cell. 
Spontaneous emission of circularly polarized light from radiative decay is a detectable signature of the vorticity of excitons. 
We propose that the exciton vorticity can also be experimentally measured by the nonlinear anomalous Hall effect, whereas the exciton superfluidity can be detected by voltage-drop quantization through a combination of the quantum geometry and the Aharonov-Casher effect. 
Topological excitons and their superfluid phase could be realized in flat bands of twisted Van der Waals heterostructures. 
\end{abstract}
\maketitle
\onecolumngrid
\newpage

\section{Introduction}
Topological phases that do not depend on protected symmetries can be traced back to the concept of quantum geometry first introduced by Michael Berry, and then developed to explain the quantization of the Hall conductivity and of various quantum Hall effects (QHE) ~\cite{Thouless1982,Haldane1988,Mele2005,Berry1984,Provost1980,Anandan1990}. This framework explains the way properties of solids can be connected to the continuous evolution in momentum of the electronic wavefunctions. It can be formalized through the idea of parallel transport of quantum states, in which periodic wavefunctions are adiabatically cycled through a variable in parameter space. In going through the adiabatic cycle, the wavefunctions pick up phases that reflect on the quantum geometry of the bands~\cite{Vanderbilt}. The non-trivial topology of Bloch wave functions is universally captured by a quantized Chern number, and the systems that display it in two dimensions (2D) are commonly termed Chern insulators~\cite{Haldane1988,Chang2013}.

No continuous symmetry is spontaneously broken in these insulators and hence no gapless collective excitation is expected in their bulk. However, electrons can be excited  from the valence band into the conduction band through the absorption of photons. Coulomb interactions  form midgap excitons, bound states comprised of excited electrons in the conduction band and holes left in the valence band, which dominate the optical response of the insulators~\cite{Haug1993}. Excitonic effects are expected to be strong in 2D because of reduced screening and enhanced electron-hole overlaps as a consequence of geometric confinement~\cite{Wang2018}.

Electron-hole (e-h) excitations in 2D Chern insulators can inherit a Berry phase from the underlying valence (v) and conduction (c) bands shown in Fig.~\ref{fig-1}(a). In the reciprocal space with states $|i,\bk \rangle$, with $i \in \{1,2\}$ a generic two-orbit index, the valence- and conduction-band states can be written in terms of the orbital basis as  $|a,\bk \rangle = \sum_{i} U_{i,a,\bk} | i,\bk\rangle$, where $a \in \{\mathrm{v,c}\}$ and $U_{a,i,\bk}$ is the periodic part of the Bloch wave function. In the electron basis, where creating a hole in the valence band with momentum $-\bk$ is equivalent to the annihilation of an electron in the same band with momentum $\bk$, the e-h pair state is defined as $|\mathrm{P},\bk \rangle =  |\mathrm{c},\bk\rangle  |\mathrm{v},\bk \rangle^*$, where  
``$*$'' denotes complex conjugation. The Berry connection of the pair state reads $\mathbf{A}_{\mathrm{P},\bk} \equiv -i \langle \mathrm{P}, \bk | \d_\bk | \mathrm{P},  \bk \rangle = \mathbf{A}_{\mathrm{v},\bk} - \mathbf{A}_{\mathrm{c},\bk}$ where $\mathbf{A}_{a,\bk} = -i \sum_{i} U^\ast_{i,a,\bk}\d_\bk U_{i,a,\bk}$ is the Berry connection of band $a$.  The Chern number of the pair state reads $\mathcal{C}_{\mathrm{P}} =  \frac{1}{2\pi}\int_\mathrm{BZ} \mathrm{d}^2 \mathbf{k} \cdot \nabla_{\bk} \times \mathbf{A}_{\mathrm{P},\mathbf{k}}= \mathcal{C}_{\mathrm{v}} - \mathcal{C}_{\mathrm{c}}$, with BZ the Brillouin zone. Therefore, if the two bands have different Chern numbers, the e-h pair state has a finite Chern number. Based on the same type of argument, it has been proposed that in superconductors with electron pairing between two disconnected Fermi surfaces, the Cooper pair state inherits a nontrivial geometric phase from the underlying Fermi surfaces~\cite{Nagaosa2003,LiHaldane2018}. 

The information about the topological nature of excitons is encoded in their profile wavefunction. Excitons are linear superpositions of electron-hole pair states, $|X,\mathbf{Q}\rangle = \sum_{\bk \in \mathrm{BZ}} \mathcal{R}_\bk (\mathbf{Q}) |e, \mathbf{k}+\mathbf{Q}\rangle |h,-\mathbf{k}\rangle$ forming an envelope function $\mathcal{R}_\mathbf{k}(\mathbf{Q})$ with a finite center of mass (COM) momentum $\mathbf{Q}$ even in centrosymmetric materials.  Excitonic bands are defined by the midgap energy dispersion of excitons with respect to their COM momentum. In conventional semiconductors, excitons can be very extended real space objects, permitting a simplification ~\cite{Haug1993,Wang2018} in which their effective mass is approximated by a profile wavefunction with zero COM momentum. In flat bands, where the effective mass is infinite and the mobility of excitons is governed by the quantum geometry of the bands, the size of the excitons is of the order of a lattice constant and their envelope wavefunction extends over COM momenta in the whole BZ. 

Topological flat bands hosting quantum anomalous Hall states have been experimentally observed in twisted bilayer graphene~\cite{Sharpe, Serlin}, which offer a valuable opportunity for the exploration of topological and quantum geometric effects.
Non-hydrogenic exciton states arising from the quantum geometry have been predicted in theory~\cite{Bychkov1980,Kallin1984,Srivastava2015,Wu2015,Zhou2015,Wang2019-1} and observed in transition metal dichalcogenide (TMD) monolayers~\cite{Beaulieu2023}. The effect of the quantum geometry on the COM motion of excitons has also been theoretically discussed~\cite{Yao2008,Kwan2021,Tang2023}. The onset of superconductivity in a partial filled flat band has been investigated theoretically and it is found that the superfluid stiffness arises purely from the quantum metric of the band~\cite{Volovik2011,PeottaTorma2015,Huhtinen2022,Jiang2023}.

In this work we theoretically show the existence of topological excitons in the midgap of generic flat Chern bands when the conduction and valence bands are topologically distinct. Although excitons can spontaneously form in narrow gap insulators~\cite{Lu,Cercellier, Bucher}, they are much easier to observe with pumped light~\cite{Koch2006}. Monochromatic photons can promote coherent population inversion between flat bands over states spanning the whole BZ. For that purpose, we develop the nonequilibrium theory for light-pumped excitons in topological flat bands and investigate the manifestations of the quantum geometry on the exciton profile functions, as well as on the exciton COM motion. We demonstrate that  the envelope wavefunction of topological excitons has a finite vorticity that is mandated by gauge invariance. We corroborate our analysis by explicitly solving the out-of-equilibrium equation of state for excitons on the lattice for the flattened Haldane model. 

Those excitons condense below the Kostelitz-Thouless transition temperature and form a novel type of topological neutral superfluid with profile wavefunctions in momentum space that carry a finite vorticity, reflecting the Chern number of the state. Exciton condensates have been observed in  double-layer systems, where  Fermi surfaces of electrons and holes are spatially separated by design~\cite{Wang2019, Li2017}, and also in the topologically trivial bands of monolayer transition-metal dichalchogenides~\cite{Kogar, Sun}. 
We propose observables that signal the nontrivial topology and quantum geometry of excitons in both insulating and superfluid phases, including the non-linear Hall effect and a realization of the Aharonov-Casher effect.
Topological excitons are chiral and expected to spontaneously emit circularly polarized light through radiative decay. Those states can be a potential platform for realizing photonic qubit gates~\cite{adami1999, Knill2001, Kok2007} and novel optoelectronic devices, such as superradiant pulse emitters~\cite{Gross1982, Scheibner2007}, in the superfluid phase.

\section{Results}

\subsection{Exciton theory in two-band insulators} 
As a minimal model, we study a two-band electron system driven by monochromatic light incident normally to the system, i.e., $\boldsymbol{E}(t) = \mathbf{E} e^{-i\W t} + \mathbf{E}^\ast e^{i \W t}$, where $\mathbf{E} = E \left( \hat{\mathbf{x}} \cos \theta + i \hat{\mathbf{y}} \sin \theta \right)$ with $E$ and $\theta$ describing the amplitude and the polarization of the light field, respectively. The corresponding vector potential can be chosen as $\mathbf{A}(t) = -\frac{i}{\W} \left( \mathbf{E} e^{-i\W t} - \mathbf{E}^\ast e^{i \W t} \right)$. 
Via the Peierls substitution in momentum space $\bk \to \bk + \frac{e}{\hbar} \mathbf{A}(t)$~\cite{Ivchenko2005,Ma2021}, with $e >0$ being the elementary charge, the system Hamiltonian is given by $\mathcal{H}_S(t) =  \sum_{\bk} \Psi_\bk^\dagger\hat{\mathcal{H}}_{\bk + \frac{e}{\hbar} \mathbf{A}(t)} \Psi_\bk  + \frac{1}{2 A} \sum_{\bq} v_\bq n_{\bq}n_{-\bq}$,
where $\Psi_\bk$ is a two-spinor in orbit space, $\hat{H}_\bk$ is the $2\times 2$ Hamiltonian matrix, $n_\bq = \sum_{\bk} \Psi^\dagger_{\bk+\bq} \Psi_{\bk}$ is the density fluctuation operator and  $v_\bq$ describes the repulsive density-density interaction. $A$ is the system area, and 
we have denoted $\sum_{\bk} \equiv \frac{A}{(2\pi)^2} \int\!d^2\bk$ for a shorthand.
We are interested in the regime where the light frequency $\W$ is near the optical resonance and the system is coupled to a reservoir in a steady state. For more details see Methods section.

Exciton states are defined in electron band basis. Diagonalizing the single-particle Hamiltonian  by a unitary matrix $\hat{U}_\bk$, i.e., $\hat{U}_\bk^\dagger \hat{\mathcal{H}}_\bk \hat{U}_\bk = \hat{\vep}_\bk \equiv \mathrm{diag}( \vep_{\mathrm{c},\bk}, \vep_{\mathrm{v},\bk} )$ with $\vep_{\mathrm{c},\bk} > \vep_{\mathrm{v},\bk}$,  one can rewrite the system Hamiltonian in band space and the dominant exciton channels as
\begin{align} \label{ham-s-eff}
\mathcal{H}_S(t) = & \sum_{\bk} \psi_\bk^\dagger \hat{\mathcal{H}}_\bk(t) \psi_\bk  \nn \\
& -\frac{1}{A} \sum_{\bq,\bq',\bQ} W_{\bq,\bq';\bQ} \Phi_{\bq+\frac{\bQ}{2},\bq-\frac{\bQ}{2}}^\dagger \Phi_{\bq'+\frac{\bQ}{2},\bq'-\frac{\bQ}{2}}.
\end{align}  
Here $\psi_\bk = \hat{U}_\bk^\dagger \Psi_\bk \equiv ( \psi_{\mathrm{c},\bk}, \psi_{\mathrm{v},\bk} )^T$ is the two-spinor in band space, where 
 $\Phi_{\bq+\frac{\bQ}{2},\bq-\frac{\bQ}{2}}^\dagger = \psi_{\mathrm{c},\bq+\frac{\bQ}{2}}^\dagger \psi_{\mathrm{v},\bq-\frac{\bQ}{2}}$ and $\Phi_{\bq+\frac{\bQ}{2},\bq-\frac{\bQ}{2}} = \psi_{\mathrm{v},\bq-\frac{\bQ}{2}}^\dagger \psi_{\mathrm{c},\bq+\frac{\bQ}{2}}$ are respectively the creation and annihilation operators of an e-h pair state with COM momentum $\bQ$ and relative momentum $\bq$.
The effective pair interaction is composed of two parts, 
$W_{\bq,\bq';\bQ} = W^{(\mathrm{d})}_{\bq,\bq';\bQ} - W^{(\mathrm{e})}_{\bq,\bq';\bQ}$,
with 
\begin{align} \label{Weff}
& W^{(\mathrm{d})}_{\bq,\bq';\bQ} = v_{\bq'-\bq} \,\mathcal{U}^{\mathrm{c}\mathrm{c}}_{\bq+\frac{\bQ}{2},\bq'+\frac{\bQ}{2}} \mathcal{U}^{\mathrm{v}\mathrm{v}}_{\bq'-\frac{\bQ}{2},\bq-\frac{\bQ}{2}}, \nn \\
& W^{(\mathrm{e})}_{\bq,\bq';\bQ} = v_\bQ \, \mathcal{U}^{\mathrm{c}\mathrm{v}}_{\bq+\frac{\bQ}{2},\bq-\frac{\bQ}{2}} \mathcal{U}^{\mathrm{v}\mathrm{c}}_{\bq'-\frac{\bQ}{2},\bq'+\frac{\bQ}{2}},
\end{align}
the direct and exchange e-h interactions, respectively~\cite{Wu2015,Wang2018}, where $\hat{\mathcal{U}}_{\bk_1,\bk_2} \equiv \hat{U}^\dagger_{\bk_1} \hat{U}_{\bk_2}$ represents the overlap of the Bloch wave functions encoding the quantum geometry of the band electrons.  For Coulomb interactions of the Rytova-Keldaysh form, that is widely used in layered semiconductors~\cite{Rytova1967,Keldysh1979}, $v_\bq \sim 1/|\bq|$ for $\bq \to 0$.

The single-particle Hamiltonian matrix coupled to light is given by $\hat{\mathcal{H}}_\bk(t) = \hat{U}_\bk^\dagger \hat{\mathcal{H}}_{\bk + \frac{e}{\hbar} \mathbf{A}(t)} \hat{U}_\bk$, which takes the multipole expansion 
\be \label{H0}
\hat{\mathcal{H}}_\bk(t) = \sum_{n=0}^\infty \frac{1}{n!} \sum_{\{\a_i\}_{i=1}^n} A_{\{\a_i\}}^{(n)}(t) \, \hat{\mathcal{T}}_{\a_n,\cdots \a_1;\bk}^{(n)},
\ee 
where $\a_i \in \{x,y\}$, and  $A_{\{\a_i\}}^{(n)}(t) = \prod_{i=1}^n [(e/\hbar) A_{\a_i}(t)] \sim O(E^n)$ represents the light-field $n$-tensor.  The electron multipole tensor $\hat{\mathcal{T}}^{(n)}$ obbeys the recursive relation $\hat{\mathcal{T}}_{a_{n+1}, a_{n}\cdots a_1;\bk}^{(n+1)} = \mathcal{D}_{\bk_{a_{n+1}}} \hat{\mathcal{T}}_{a_{n},\cdots a_1;\bk}^{(n)}$ with  $\hat{\mathcal{T}}_\bk^{(0)} = \hat{\vep}_\bk$ and the differential superoperator $\mathcal{D}_{\bk} = \mathcal{\d}_{\bk} + i[\hat{\boldsymbol{\mathcal{A}}}_\bk, \quad ]$, in which $\hat{\boldsymbol{\mathcal{A}}}_\bk = -i\hat{U}_\bk^\dagger \d_{\bk} \hat{U}_\bk$ is the Berry connection matrix. We note that in general $A_{\{\a_i\}}^{(n)}(t)$ involves only $m_\mathrm{th}$ order harmonics, with $m \in \{-n,-n+2,\cdots,n\}$. 

Up to linear order of the electric field, which captures the single-photon processes, the driven band Hamiltonian \eqref{H0} takes the simple form 
\begin{align} \label{H-driven}
& \hat{\mathcal{H}}_\bk(t) = 
\begin{pmatrix} 
\vep_{\mathrm{c},\bk} & e^{-i \W t} \, \mathbf{E} \cdot \mathbf{P}_{\mathrm{cv},\bk} \\ 
 e^{i \W t} \, \mathbf{E}^\ast \cdot \mathbf{P}_{\mathrm{cv},\bk}^\ast & \vep_{\mathrm{v},\bk} \end{pmatrix}, \nn \\
 & \mathbf{P}_{\mathrm{cv},\bk} = \frac{e \Delta_{\bk}}{\W} \boldsymbol{\mathcal{A}}_{\mathrm{cv},\bk},
\end{align}
with $\Delta_{\bk} = \vep_{\mathrm{c},\bk} - \vep_{\mathrm{v}\bk}$ being the direct gap between conduction and valance bands and $\mathbf{P}_{\mathrm{cv},\bk}$ is the effective interband electric dipole moment arising from the interband Berry connection $\boldsymbol{\mathcal{A}}_{\mathrm{cv},\bk}$.  In Eq.~\eqref{H-driven} we have kept only the light-field terms relevant to optical resonances \cite{Morimoto2016-1} (for details see Supplementary Information). The upper bound on the magnitude of the interband dipole moment can be estimated from dimensional analysis. For perfect flat bands in the BZ, where the lattice constant $a_0$ is the only available length scale,  $P_{\mathrm{cv},\bk} \lesssim e a_0$.  For gapped Dirac fermions with gap $\Delta _0$ and Fermi velocity  $v_F$, one finds that $P_{\mathrm{cv},\bk} \sim e \hbar v_F/\Delta_0$.

We investigate the dynamics of the driven system exploiting the Keldysh formalism~\cite{Kamenev2011} and introduce the e-h pair order parameter $X_{\bQ,\bq}(t)$ via the Hubbard-Stratonovich transformation. In analogy to superconductivity, the order parameter $X_{\bQ,\bq}(t)$ is a dynamical pairing potential of an electron and a hole. 
After integrating out the fermions, we obtain the effective action for the order parameter as well as the generating function of the e-h pair correlation functions. The detailed Keldysh formulation including the equation of motion of $X_{\bQ,\bq}(t)$ is presented in Methods. We note that the light field only couples to the relative motion of the electron and hole since the photons do not carry in-plane momentum. Moreover, up to linear order of the electric field, the order parameter only involves the first harmonic. Eventually, we obtain the order parameter $X_{\bQ,\bq}(t) = \delta_{\bQ,0}  \mathsf{X}_{\bq}(\W) \, e^{-i\W t}$, with the coefficient 
\begin{align} \label{X-q}
& \mathsf{X}_{\bq}(\W) = \mathbf{E} \cdot \boldsymbol{\mathcal{P}}_\bq, \nn \\
& \boldsymbol{\mathcal{P}}_\bq = \sum_{\bq',\bq''} \Pi^{R}_{\bq,\bq'}(\W,\bQ=0) \, W_{\bq',\bq'';\bQ=0} \mathbf{P}_{\mathrm{cv},\bq''},
\end{align}
where $\Pi^{R}_{\bq,\bq'}(\w,\bQ)$ is the retarded propagator of an e-h pair, which defines the effective Hamiltonian of an e-h excitation  incorporating the Coulomb interaction (see Methods), and $\boldsymbol{\mathcal{P}}_\bq$ is the electric polarization of the e-h pair, which is a linear combination of the band dipole moments averaged over the whole BZ. Equation~\eqref{X-q} is consistent with the theory of exciton-light interaction within the dipole approximation~\cite{Haug1993,Ivchenko2005}. 

The poles of the e-h pair retarded propagator in Eq.~\eqref{X-q} define the exciton Hamiltonian
\be \label{ham-ex}
h_{\bq,\bq'}(\bQ) = \del_{\bq,\bq'}  \Delta_{\bq+\frac{\bQ}{2},\bq'-\frac{\bQ}{2}} - Z_{\bq,\bq';\bQ} W_{\bq,\bq';\bQ}, 
\ee  
where $\Delta_{\bk_1,\bk_2} = \vep_{\mathrm{c},\bk_1} - \vep_{\mathrm{v}\bk_2}$ is the indirect band gap, and $Z_{\bq,\bq';\bQ} = \sqrt{\del{f}_{\bq+\bQ/2,\bq-\bQ/2} \del{f}_{\bq'+\bQ/2,\bq'-\bQ/2}}$  is the spectrum weight that varies from zero to one, with $\del{f}_{\bk_1,\bk_2} = f_{F}(\ep_{\mathrm{v},\bk_1})-f_{F}(\ep_{\mathrm{c},\bk_2})$ and $f_{F}$ the equilibrium Fermi distribution. The spectrum weight is determined by the temperature $T$ and chemical potential $\mu$ of the reservoir electrons, is maximized when $\mu$ is at the mid gap, and can be significantly suppressed when $T$ is comparable to the band gap. 
The eigenvalue problem of Hamiltonian \eqref{ham-ex} gives the e-h pair eigenstates,
\be \label{wannier}
\sum_{\bq'} h_{\bq,\bq'}(\bQ) \mathcal{R}_{\nu,\bq'} (\bQ) = \vep_{\nu}(\bQ) \mathcal{R}_{\nu,\bq} (\bQ),  
\ee
where $\mathcal{R}_{\nu,\bq}(\bQ)$ is the profile wavefunction of the pair states  with energy $\vep_\nu(\bQ)$, and $\nu$ labels a complete set of quantum numbers obeying the orthonormalization condition $\sum_{\bq \in \mathrm{BZ}} \mathcal{R}_{\nu,\bq}^\ast(\bQ) \mathcal{R}_{\nu',\bq}(\bQ) =\del_{\nu\nu'}$. The eigenstates with discrete midgap energies are defined as the exciton bound states with corresponding  exciton bands $\vep_\nu(\bQ)$. We note that Eq.~\eqref{wannier} is the exciton Wannier equation~\cite{Haug1993} in reciprocal space. 

The e-h pair retarded propagator in Eq.~\eqref{X-q} has the spectral representation 
\be
\Pi^{R}_{\bq,\bq'}(\w,\bQ) = Z_{\bq,\bq';\bQ} \sum_\nu \frac{\mathcal{R}_{\nu,\bq}(\bQ) \mathcal{R}_{\nu,\bq'}^\ast(\bQ)}{\w - \vep_\nu(\bQ) + i \G},
\ee
where $\G$ is the system-to-reservoir electron tunneling rate. Since the effective e-h polarization in Eq.~\eqref{X-q} only couples the exciton states at $\bQ = 0$, we define $\vep_\nu \equiv \vep_\nu(\bQ=0)$ and $\mathcal{R}_{\nu,\mathbf{q}} \equiv \mathcal{R}_{\nu,\mathbf{q}}(\bQ=0)$ as shorthand notations. In perfectly flat bands, the band gap is a constant $\Delta_{\bk_1,\bk_2} = \Delta$ and the polarization in Eq.~\eqref{X-q} can be written as
\be \label{ex-dipole}
\boldsymbol{\mathcal{P}}_\bq = \sum_{\nu} \frac{\vep_{\mathrm{B},\nu}\boldsymbol{\ell}_\nu \mathcal{R}_{\nu,\bq}}{\W - \vep_\nu + i \G}, \quad \boldsymbol{\ell}_\nu = \sum_{\bq} \mathcal{R}^\ast_{\nu,\bq} \mathbf{P}_{\mathrm{cv},\bq}, 
\ee 
where 
$\vep_{\mathrm{B},\nu} \equiv \Delta - \vep_\nu$ is the binding energy. $\boldsymbol{\ell}_\nu$ is the effective exciton dipole moment of state $\nu$ that determines the optical selection rule, i.e., bound state $\nu$ couples to the light field only if $\mathbf{E}\cdot \boldsymbol{\ell}_\nu \neq 0$.   
    The time-dependent mean-field Hamiltonian of the quasiparticles reads 
\begin{align} \label{TDMF}
\hat{\underline{\mathcal{H}}}_{\bk}(t) = & \hat{\mathcal{H}}_{\bk}(t) + \begin{pmatrix} 0 &  \mathsf{X}_{\bk}(\W) e^{-i\W t} \\ \mathsf{X}_{\bk}^\ast(\W) e^{i\W t} & 0 \end{pmatrix}  \nn \\
= & \begin{pmatrix} 
\vep_{\mathrm{c},\bk} & e^{-i \W t} \, \mathbf{E} \cdot \underline{\mathbf{P}}_{\mathrm{cv},\bk} \\ 
 e^{i \W t} \, \mathbf{E}^\ast \cdot \underline{\mathbf{P}}_{\mathrm{cv},\bk}^\ast & \vep_{\mathrm{v},\bk} \end{pmatrix},
\end{align} 
where $\underline{\mathbf{P}}_{\mathrm{cv},\bk} = \mathbf{P}_{\mathrm{cv},\bk} + \boldsymbol{\mathcal{P}}_\bk$ is the effective interband electric dipole moment that incorporates the light-pumped exciton order parameters in Eq.~\eqref{X-q}. Equations \eqref{X-q} and \eqref{TDMF} form the complete theory for excitons in a two-band flat-band insulator excited by one-photon processes. 

Since the excitons contribute to the dielectric polarization in addition to the noninteracting e-h pairs, the exciton levels $\vep_\nu$ can be observed from the resonance peaks of the reflection coefficient~\cite{Haug1993}. At normal light incidence and near resonance, $\W \sim \vep_\nu$, the reflection coefficient has the standard form $r(\W) = i \G_{0,\nu}/[\W-\vep_\nu+i(\G_{0,\nu}+\G)]$, where $\G_{0,\nu} = \frac{4\pi^2 \del{f} |\boldsymbol{\ell}_\nu|^2}{\hbar \l \ep A_0}$ is the radiative decay rate, with $A_\mathrm{0}$ the unit cell area, $\l = 2\pi c/\W$ the light wavelength in the insulator and $\ep = \ep_\mathrm{r} \ep_0$  the high-frequency dielectric constant~\cite{Ivchenko2005}. 
Dimension analysis gives an upper bound for the exciton dipole moment in flat bands, $\ell_\nu \lesssim e a_0$, with $a_0$  the lattice constant. Hence,  $\G_{0,\nu} \lesssim \vep_{\nu} \a_0/(\hbar \sqrt{\ep_\mathrm{r}})$ where $\a_0 \approx 1/137$ is the fine structure constant. For $\vep_{\nu} \sim 1 \,\mathrm{eV}$ and $\ep_{r} \sim 10$ as in TMD monolayers, the estimated exciton radiative lifetime $1/ \G_{0,\nu}$ is longer than  hundreds of femtoseconds, in agreement with first principle calculations~\cite{Palummo2015,WangH2016}. In twisted bilayer systems like graphene, where small gap flat bands are usually observed, much longer radiative lifetimes are expected.

\subsection{Exciton bound states with nontrivial vorticity}
We first examine  the topology of the exciton states implied by charge U(1) gauge symmetry. Applying local U(1) gauge transformations to the conduction and valence electron wave functions $\hat{U}_\bk \to \hat{U}_\bk \mathrm{diag}\{ e^{i\phi_{\mathrm{c},\bk}}, e^{i\phi_{\mathrm{v},\bk}}\}$, where $\phi_{a,\bk}$ with $a \in\{\mathrm{c,v}\}$ are arbitrary phase fields in reciprocal space, the exciton Hamiltonian transforms as $h_{\bq,\bq'}(\bQ)  \to e^{i (\phi_{\mathrm{c},\bq'+\bQ/2}-\phi_{\mathrm{c},\bq+\bQ/2})} e^{i (\phi_{\mathrm{v},\bq-\bQ/2}-\phi_{\mathrm{v},\bq'-\bQ/2})} h_{\bq,\bq'}(\bQ)$. 
In order to keep the Wannier equation \eqref{wannier} gauge-invariant, one should define the U(1) gauge transformation of the exciton wavefunction as 
\be \label{u1-ex}
\mathcal{R}_{\nu,\bq}(\bQ) \to e^{-i(\phi_{\mathrm{c},\bq+\bQ/2} - \phi_{\mathrm{v},\bq-\bQ/2})} \mathcal{R}_{\nu,\bq}(\bQ).
\ee
To incorporate the geometric phase of the electron wave functions we take $\phi_{a,\bk} = \int_{\bk_0}^{\bk} d\bk' \cdot \boldsymbol{\mathcal{A}}_{a,\bk'}$, where 
$\boldsymbol{\mathcal{A}}_{a,\bk}$ is the intraband Berry connection and the initial point $\bk_0$ and the integral path from $\bk_0$ to $\bk$ are gauge degrees of freedom. Choosing an exciton wave function $\mathcal{R}_{\nu,\bq}(\bQ) \equiv |\mathcal{R}_{\nu,\bq}(\bQ)| e^{i \varphi_{\nu,\bq}(\bQ)}$ with the phase field analytic in the whole reciprocal space, we introduce the geometric phase of the exciton state $\tilde{\varphi}_{\nu,\bq}(\bQ) \equiv \varphi_{\nu,\bq}(\bQ) + \int_{\bk_0}^{\bq+\bQ/2} d\bk' \cdot \boldsymbol{\mathcal{A}}_{\mathrm{c},\bk'} - \int_{\bk_0}^{\bq-\bQ/2} d\bk' \cdot \boldsymbol{\mathcal{A}}_{\mathrm{v},\bk'}$. Therefore, the phase gradient field $\mathbf{u}_\bq(\bQ) \equiv \d_\bq\tilde{\varphi}_{\nu,\bq}(\bQ) = \d_\bq \varphi_{\nu,\bq}(\bQ) - \boldsymbol{\mathcal{A}}_{\mathrm{c},\bq+\bQ/2} + \boldsymbol{\mathcal{A}}_{\mathrm{v},\bq-\bQ/2}$ is a gauge-invariant observable. The corresponding vorticity is defined as $\boldsymbol{\zeta}_{\bq}(\bQ) \equiv 
\boldsymbol{\nabla}_\bq \times \mathbf{u}_\bq(\bQ) = -\boldsymbol{\W}_{\mathrm{c},\bq+\bQ/2}+\boldsymbol{\W}_{\mathrm{v},\bq-\bQ/2} = \zeta_{\bq}(\bQ) \hat{\mathbf{z}}$, where $\hat{\mathbf{z}}$ is the unit vector in $z$ direction and $\boldsymbol{\W}_{a,\bk} = \boldsymbol{\nabla}_\bk \times \boldsymbol{\mathcal{A}}_{a,\bk}$ is the $a$-band Berry curvature. We obtain quantized total vorticity
\be
\zeta = \int_\mathrm{BZ} \! \frac{d^2\bq}{2\pi} \zeta_{\bq}(\bQ) = \mathcal{C}_\mathrm{v} - \mathcal{C}_\mathrm{c},
\ee
which is independent of the COM momentum $\bQ$. For $\mathcal{C}_\mathrm{c} = - \mathcal{C}_\mathrm{v}$, we obtain exciton wavefunctions with vorticity that is twice of the valence-band Chern number $\nu = 2 \mathcal{C}_\mathrm{v}$. Note that this analysis is valid for excitons in general, not only for those in flat bands. 

One can readily see formally why  the topology of the electron wavefunctions is inherited by the exciton wavefunctions. For a non-vanishing Chern number, 
we could choose a gauge in which the electron wave function is analytic anywhere in the BZ except at isolated points~\cite{Dirac1931}, or else divide the BZ into distinct analytic regions separated by non-analytic boundaries~\cite{WuYang1975}. This non-analyticity  is incorporated in the exciton Hamiltonian \eqref{ham-ex} through wavefunction overlaps $\hat{\mathcal{U}}$ that appear in the effective interaction in Eq.~\eqref{Weff}. The conventional group-theoretical analysis of exciton wavefunctions does not apply to topological excitons, since the wavefunctions of the latter are not analytic~\cite{Nagaosa2003,LiHaldane2018}. Due to the emergence of gauge degrees of freedom of the exciton states in Eq.~\eqref{u1-ex}, the pairing ``symmetry'' is gauge-dependent, even though the vorticity is invariant.

In order to show an explicit example of the vortex exciton states we study a simple model of two flat bands, $\mathcal{H}_\bk = \frac{\Delta}{2} \hat{\mathbf{d}}(\bk) \cdot \boldsymbol{\s}$, where $\hat{\mathbf{d}}(\bk)$ is a 3D unit vector. We parametrize the $\hat{\mathbf{d}}(\bk)$ vector by an arbitrary meromorphic function $\chi(z)$ of the complex variable $z \equiv k_x + i k_y$ as $\hat{d}_1 + i \hat{d}_2 = \frac{2 \chi(z)}{1+|\chi(z)|^2}$, $\hat{d}_1 - i \hat{d}_2 = \frac{2 \chi^\ast(z)}{1+|\chi(z)|^2}$ and $\hat{d}_3 = \frac{1-|\chi(z)|^2}{1+|\chi(z)|^2}$~\cite{Jian2013}. It is easy to obtain the electron eigenstates 
$\hat{U}_\bk = \frac{1}{\sqrt{1+|\chi|^2}} \begin{pmatrix} 1 & \chi^\ast \\ \chi & -1 \end{pmatrix}$ up to a U(1) gauge transformation. One can prove that the Chern number of the conduction band is equal to the number of poles of $\chi(z)$, with ``Dirac strings'' along the third extra dimension piercing the BZ. 
We can simply study the $\bQ=0$ component of the Wannier equation since the total vorticity is independent of the COM motion. The effective interaction reads
$W_{\bk,\bk^\prime;\bQ=0} = v_{\bk'-\bk} \frac{ \left( 1 +  \chi^\ast(z) \chi(z') \right)^2 }{(1+|\chi(z)|^2)(1+|\chi(z')|^2)}$. 
Assuming a single pole at $\bk=0$, one has $\chi(z) \approx a/z$ in the vicinity of $\bk=0$, where $a$ is the residue of the pole characterizing the pole strength and the c-band Berry curvature is $\W_\mathrm{c} \approx 4\pi/|a|^{2}$. For $z,z'$ in the vicinity of $z=0$ and up to quadratic order in $\bk,\bk'$, we obtain
\be \label{U-poles}
W_{\bk,\bk^\prime;\bQ=0} \approx v_{\bk'-\bk} e^{2i (\phi-\phi')} \l_{\bk,\bk'},
\ee    
where $\l_{\bk,\bk'} = 1 - \frac{\W_\mathrm{c}}{4\pi} [ |\bk-\bk'|^2 - 2 i (\bk \times \bk')_z ]$ and we have defined the polar coordinates $\bk = k(\cos\phi,\sin\phi)$. 

The correction from the quantum geometry to the bare interaction in equation \eqref{U-poles} is composed of two parts. The first one is the topological factor $e^{2i (\phi-\phi')}$, that incorporates the pole of $\chi(z)$. This term modifies the exciton wave function but does not influence the energy spectrum. The presence of the poles (nontrivial Chern number) is not a perturbation and the Taylor expansion of $\chi(z')$ about $z$ cannot capture the pole at $z=0$. The second one is the geometric factor $\l_{\bk,\bk'}$, that is perturbative in the Berry curvature $\W_\mathrm{c}$,  which modifies the exciton energy spectrum. 
The vorticity is incorporated in the topological factor. We assume that for $e^{2i (\phi-\phi')}=1$ the Wannier equation~\eqref{wannier} at $\bQ=0$ takes an eigenstate $\mathcal{R}_{\nu,\bk} = |\mathcal{R}_{\nu,\bk}|e^{i \varphi_{\nu,\bk}}$, that is an analytic function. Absorbing the topological factor into the eigenstate, the wave functions can be written as $\mathcal{R}_{\nu,\bk} e^{2i \phi}$, so that vorticity is the winding number $\zeta = 2$. Equation \eqref{U-poles} is the effective interaction in a ``symmetric'' gauge, that is, $\l_{\bk,\bk'}$ is invariant under a global rotation about $\bk=0$. In the limit of $\W_\mathrm{c} \to 0$, the ground state wave function should take a $d+id$-wave form $\mathcal{R}_{0,\bk} = \mathcal{R}_{0,k} e^{i 2 \phi}$ for conventional Coulomb interaction. 

\begin{figure}[h]
\centering
\includegraphics[trim=0cm 0cm 0cm 0cm, clip=true, width=0.9\textwidth]{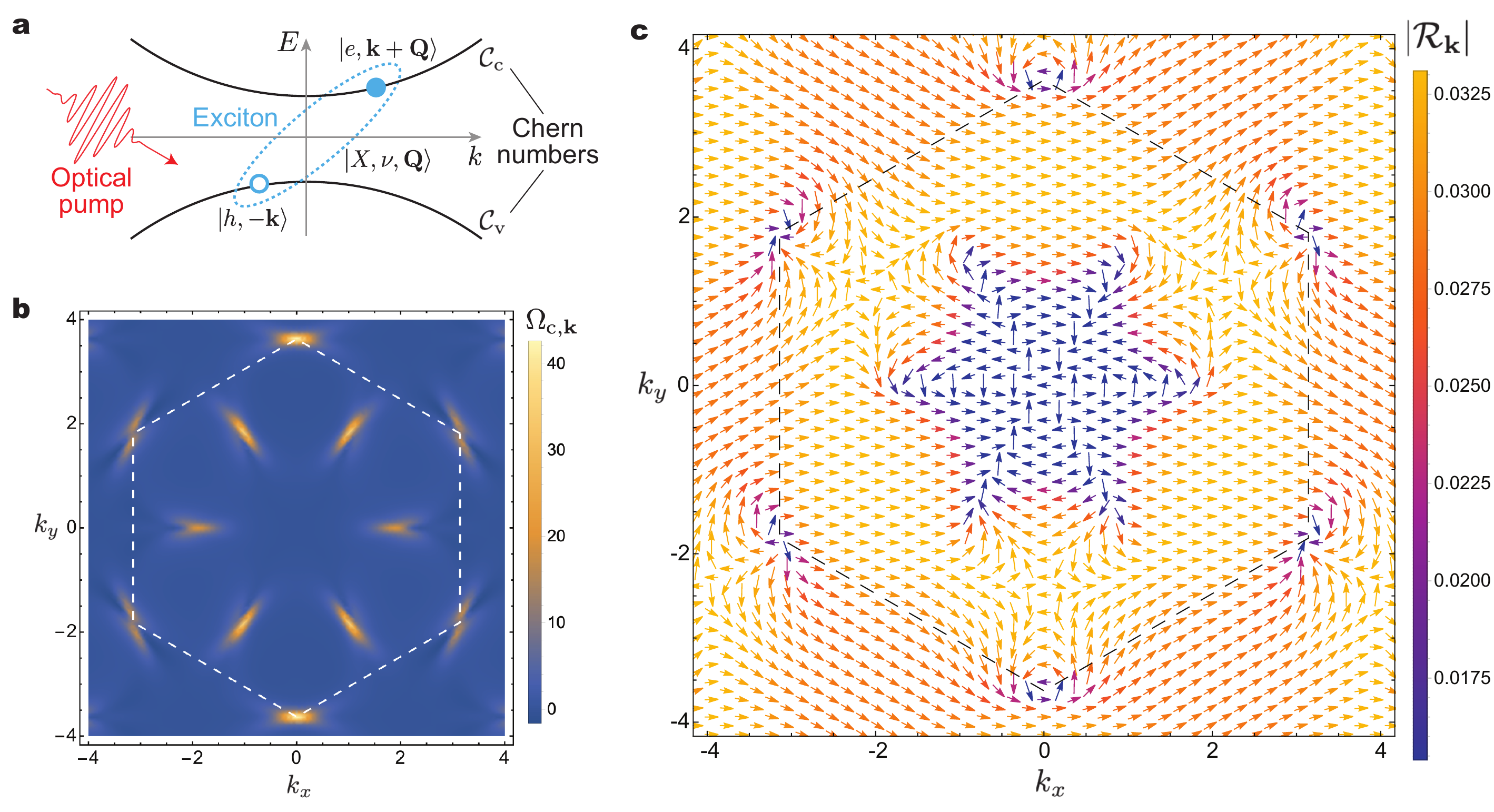}
\caption{
\textbf{Exciton bound state in a flat-band Chern insulator.} 
\textbf{(a)} Schematics of an exciton excitation in a two-band insulator. Electrons can be excited from the valence band (VB) into the conduction band (CB) through optical processes. Coulomb interaction binds an excited electron in CB and state $|e,\bk+\bQ\rangle$ with a hole left in VB at state $|h,-\bk\rangle$, forming midgap exciton states $|X,\nu,\bQ\rangle$. $\nu$ denotes a set of quantum of numbers describing the exciton bound states and $\bQ$ is the COM momentum. The excitons inherit a Berry phase from the underlying bands when the latter have different Chern numbers, $\mathcal{C}_\mathrm{v}$ and $\mathcal{C}_\mathrm{c}$. Quantum geometric effects are dominant in the limit of flat bands. 
\textbf{(b)} CB Berry curvature $\W_{\mathrm{c},\bk}=-\W_\mathrm{v,\bk}$ of the flattened Haldane model (see Methods). We take the lattice constant $a=1$ and the first BZ is indicated by dashed lines.
\textbf{(c)} Momentum-space exciton wave function at rest COM motion calculated from the flattened Haldane model. The black dashed lines indicate the first Brillouin zone boundary. At each $\bk$ point the color and orientation of the arrow represent the amplitude and phase of the exciton wavefunction $\mathcal{R}_\bk$.} 
\label{fig-1}
\end{figure}

For concreteness, we explicitly verify this analysis by numerically solving the eigenvalue equation~\eqref{wannier} on the lattice for the flattened Haldane model in the presence of a contact interaction, as detailed in Methods. This model has two Chern bands with $\mathcal{C}_\mathrm{v}=-\mathcal{C}_\mathrm{c}=1$. 
Fig.~\ref{fig-1}(b) shows the Berry curvature distribution of the model in the conduction band.  The solution of the exciton eigenfunction $\mathcal{R}_{\bk}$  is shown in Fig.~\ref{fig-1}(c), and has an approximate $d+id$ symmetry around the center of the BZ. We note that while the symmetry of the exciton wave function depends on microscopic details, the vorticity of the solution ($\zeta=2$) is completely fixed by the topology.  

\subsection{Effective exciton masses and superfluidity}
\begin{figure}[h]
\centering
\includegraphics[trim=0cm 0cm 0cm 0cm, clip=true, width=0.9\textwidth]{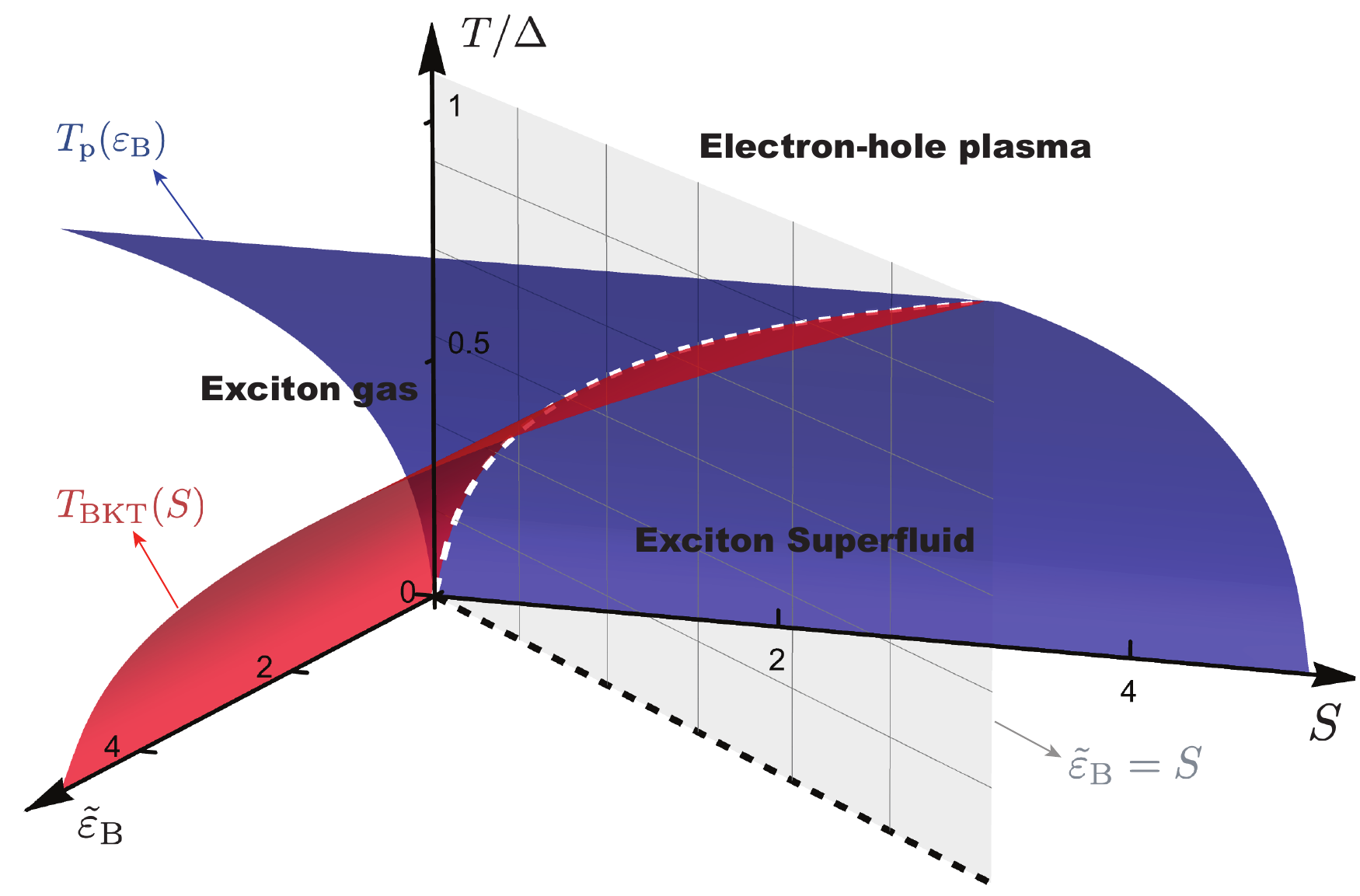}
\caption{
\textbf{Phase diagram of excitons as a function of temperature $T$, zero-temperature binding energy $\td{\vep}_\mathrm{B}$, and effective light intensity $S$.} Energies are normalized by the band gap $\D$, with $\td{\vep}_\mathrm{B} \equiv \vep_\mathrm{B}(0)/\D$.
 The pairing temperature $T_\mathrm{p}$ is determined by the binding energy $\vep_\mathrm{B}(0)$ (blue surface). As temperature  crosses $T_\mathrm{p}$ from above, exciton bound states form among light-pumped electron-hole plasma. The superfluid transition temperature $T_\mathrm{BKT}$ is a function of the effective intensity of the light pump $S$ (red surface). $T_\mathrm{BKT}$ is determined by the superfluid density in Eq.~\eqref{stiffness}. The finite effective mass of excitons arises from the quantum metric of the underlying electron bands as well as from the Coulomb interaction. When $T <  T_\mathrm{BKT}$, global phase coherence develops among excitons, which condense. We note that $T_\mathrm{p} = T_\mathrm{BKT}$ for $\td{\vep}_\mathrm{B} = S$ (gray surface). } 
\label{fig-2}
\end{figure}

The COM motion of excitons in flat bands originates from the effective e-h interaction in the Wannier equation \eqref{wannier}. We rewrite the exciton Hamiltonian \eqref{ham-ex} in the $\bQ=0$ exciton basis, $\td{h}_{\mathrm{ex},\nu\nu'}(\bQ) = \sum_{\bq,\bq'} \mathcal{R}_{\nu,\bq}^\ast h_{\mathrm{ex},\bq\bq'}(\bQ) \mathcal{R}_{\nu',\bq'}$, and up to quadratic order in the COM momentum $\bQ$, we obtain
\begin{align}
\td{h}_{\mathrm{ex},\nu\nu'}(\bQ) = & \del_{\nu\nu'} \vep_{\nu} + \del{f}\sum_{\a=x,y} \mathcal{V}_{\a,\nu\nu'} Q_\a \nn \\
& + \frac{\del{f}}{2} \sum_{\a,\b=x,y} \mathcal{M}_{\a\b,\nu\nu'}^{-1} Q_\a Q_\b,
\end{align}
where the velocity tensor $\mathcal{V}_{\a,\mu\mu'}$ and the mass tensor $\mathcal{M}_{\a\b,\nu\nu'}^{-1}$ are proportional to the interaction amplitude and the differential properties of the single-particle wave functions (see Supplementary Information).  For exciton state $\nu$, the pairing temperature can be defined by the binding energy through the relation $T_\mathrm{p} = \vep_{\mathrm{B},\nu}(T_\mathrm{p})$. For flat bands, $\vep_{\mathrm{B},\nu}(T) = \del{f}(T) \vep_{\mathrm{B},\nu}(0)$. The pairing temperature as a function of zero-temperature binding energy is shown in the blue surface in Fig.~\ref{fig-2}. In inversion-symmetric systems, $\mathcal{V}_{\a,\nu\nu'} =0$, the effective inverse mass tensor of the exciton state $\nu$ is $\mathcal{M}_{\a\b,\nu\nu}^{-1}$. If the light frequency is resonant with the exciton state $\nu$, the superfluid density tensor is
\be \label{stiffness}
\rho_{\nu,\a\b} = \frac{\del{f} |\mathbf{E} \cdot \boldsymbol{\ell}_\nu|^2}{A_0 \G^2} \mathcal{M}_{\a\b,\nu\nu}^{-1},   
\ee   
where $A_0$ is the unit cell area. Therefore, the superfluid transition should occur at the Kosterlitz-Thouless temperature $T_\mathrm{BKT} = \frac{\hbar^2 \pi}{8} \sqrt{\mathrm{det} \hat{\rho}_\nu(T_\mathrm{BKT})}$, which is shown  in the red surface in Fig.~\ref{fig-2} as a function of the effective light intensity $S \equiv \frac{\hbar^2 \pi |\mathbf{E} \cdot \boldsymbol{\ell}_\nu|^2}{8 A_0 \G^2 \Delta} \sqrt{\mathrm{det} \hat{\mathcal{M}}_{\nu\nu}^{-1}}$. For a fixed light intensity, the system undergoes three phases as temperature decreases, i.e. electron-hole plasma, exciton gas and exciton superfluid. 

\subsection{Nonlinear Hall conductivity}
Photocurrents can be induced in insulators by pumping light. Application of a static electric field $\boldsymbol{\mathcal{E}}$ in the presence of pumped light induces an electric current in addition to the photocurrent. 
There are three current terms up to third order in electric fields, 
 \be
 J^\a =\s_{0,\a\b} \mathcal{E}_\b + \Sigma_{\a\b\g} E_\b E_\g^\ast +  \chi_{\a\b\g\del} E_\b E_\g^\ast \mathcal{E}_\del,
 \ee
 where $\s_{0,\a\b}$, $\Sigma_{\a\b\g}$, and $\chi_{\a\b\g\del}$ are the dc conductivity~\cite{Thouless1982}, the photoconductivity~\cite{vBK1981,Hosur2011,McIver2012,Jotzu2014,Sentef2015,Morimoto2016-1,Morimoto2016-2,Pedersen2015,Chan2021}, and the non-linear conductivity induced by light~\cite{Morimoto2016-1,Morimoto2016-2}, respectively. Anomalous Hall conductivities reflect the total Berry fluxes over the BZ. For  Hamiltonian \eqref{TDMF}, the anomalous Hall conductivity takes the expression $\s_\mathrm{NLH} = \frac{e^2}{\hbar} \frac{\del f}{(\W-\Delta)^2 + \G^2} \int_\mathrm{BZ}\frac{d^2\bk}{(2\pi)^2} |\mathbf{E} \cdot \underline{\mathbf{P}}_{\mathrm{cv},\bk}|^2 \zeta_\bk$, where $\zeta_\bk = \nabla_\bk \times (\d_\bk \varphi_\bk + \boldsymbol{\mathcal{A}}_{\mathrm{v},\bk} - \boldsymbol{\mathcal{A}}_{\mathrm{c},\bk})$ is the vorticity of e-h pairs, with $\varphi_\bk = \mathrm{Arg}(\mathbf{E} \cdot \underline{\mathbf{P}}_{\mathrm{cv},\bk})$~\cite{Morimoto2016-1,Morimoto2016-2}. We obtain 
\be \label{NLHE} 
\s_\mathrm{NLH}(\W) =\frac{e^2}{h}  \left( \sum_\nu \frac{\eta_\nu^2}{(\W-\vep_\nu)^2+\G^2} + \frac{\eta_\mathrm{cv}^2}{ (\W-\Delta)^2+\G^2} \right), 
\ee
where the weights $\eta_\nu^2 = \del f |\mathbf{E}\cdot \boldsymbol{\ell}_\nu|^2 \int_\mathrm{BZ}\frac{d^2\bk}{2\pi} |R_{\nu,\bk}|^2 \zeta_{\bk}$ and $\eta_\mathrm{cv}^2 = \del f \int_\mathrm{BZ}\frac{d^2\bk}{2\pi} |\mathbf{E}\cdot \mathbf{P}_{\mathrm{cv},\bk}|^2 \zeta_\bk$,  encode the optical selection rules and the vorticity field of e-h pairs. 

In the presence of rotational invariance, $|R_{\nu,\bk}|$ is angle-independent and the nonlinear Hall conductivity at  exciton resonant frequencies is quantized, since $\int_\mathrm{BZ}\frac{d^2\bk}{2\pi} |\mathcal{R}_{\nu,\bk}|^2 \zeta_\bk = \zeta$. One could measure the non-linear Hall conductivity in the Corbino geometry shown in Fig.~\ref{fig-3}(a). For simplicity, we assume that only one exciton bound state with energy $\vep_0$ is formed. We schematically show the non-linear Hall conductivity as a function of light frequency in Fig.~\ref{fig-3}(b). When $T > T_\mathrm{p}$, only the e-h peak with amplitude  $(\eta_\mathrm{cv}/\G)^2$ can be observed at $\W = \Delta$; when $T<T_\mathrm{p}$ an exciton peak with amplitude  $\zeta \del{f} |\mathbf{E}\cdot \boldsymbol{\ell}_\nu|^2/\G^2 $ forms at $\W = \vep_0$. The amplitude of this peak is proportional to the exciton vorticity $\zeta$. 

\subsection{Aharonov-Casher effect and electric field quantization} 
Neutral particles with a finite magnetic moment $\mathbf{m}$ couple to an external static electric field $\boldsymbol{\mathcal{E}}$ through a gauge field  $\mathbf{A} = \mu \epsilon\, \mathbf{m}\times \boldsymbol{\mathcal{E}}$ ~\cite{AC1984,Cimmino1989,Mooij1993,Molenkamp2006,Nakata2014}, where $\mu$ and $\epsilon$ are the magnetic and electric permeabilities, respectively. In the superfluid phase, phase coherence dictates that the circulation of the gauge field $\theta_\mathrm{AC} =\frac{1}{\hbar}\oint \mathrm{d} \mathbf{r} \cdot \mathbf{A} = 2 \pi N$ produces a quantized geometric phase, with $N$ an integer. This quantization is known as the Aharonov-Casher effect \cite{AC1984}, in analogy with the superconducting case. 

One could exploit the Aharonov-Casher effect to detect the superfluidity of the excitons, since they can inherit a magnetic dipole moment from the electron bands. We start from the thermodynamic definition of the orbital magnetization of the bound state $\nu$, $\mathbf{m}_\a(\bQ) \equiv -\d_\mathbf{B} \vep_\nu(\mathbf{B},\bQ)|_{\mathbf{B=0}}$, where $\vep_\nu(\mathbf{B})$ is the bound state energy in the presence of a uniform static magnetic field $\mathbf{B}$~\cite{Shi2007}. Up to leading order in the magnetic field $\mathbf{B}$, the exciton Hamiltonian reads $h_{\mathrm{ex},\bq,\bq'}'(\bQ) = h_{\mathrm{ex},\bq,\bq'}(\bQ) - \mathbf{B} \cdot \mathbf{M}_{\bq,\bq'}(\bQ)$, where $\mathbf{M}_{\bq,\bq'}(\bQ)$ are the orbital magnetization matrix element in e-h space, so that $\mathbf{m}_\nu(\bQ) = \sum_{\bq,\bq'} \mathcal{R}_{\nu,\bq}^\ast(\bQ) \mathbf{M}_{\bq,\bq'}(\bQ) \mathcal{R}_{\nu,\bq'}(\bQ)$. At zero temperature, the leading contributions to the magnetization arise from the Zeeman effect in the single-electron band and from the quantum geometry. 

For a pair of conduction and valence bands embedded in a set of flat bands, both the effective gap and the interaction term are modified by the magnetic field perturbation. At $\bQ=0$, the magnetic moment 
$m_{\nu}(\bQ=0) = \frac{1}{\hbar} \sum_\bk |\mathcal{R}_{\nu,\bk}|^2 \sum_{a=\mathrm{c,v}} \sum_{b \neq \mathrm{c,v}} [2 (\Delta-\vep_\nu)+ \s_a \Delta_{ab}] D_{ab,\bk}$, where $\s_\mathrm{c(v)}=+1(-1)$ and $D_{ab,\bk} = -e \mathrm{Im}( \mathcal{A}_{ab,x,\bk} \mathcal{A}_{ab,y,\bk}^\ast )$. For a two-band model with particle-hole symmetry, the magnetic dipole moment vanishes since the orbital magnetizations of the two bands exactly compensate each other.  For finite COM momentum $\bQ \neq 0$, both the effective gap and the interaction couple to the magnetic field, 
$m_{\nu}(\bQ) = \frac{\vep_\nu}{\hbar} \sum_\bk |\mathcal{R}_{\nu,\bk}(\bQ)|^2 \left( D_{\mathrm{cv},\bk+\bQ/2} - D_{\mathrm{cv},\bk-\bQ/2} \right)$. The derivation is presented in Supplementary Information 

In practice, for any multi-flat band system away from half filling, as in twisted bilayer graphene at 3/4 filling~\cite{Sharpe, Serlin}, the exciton magnetic moment can be finite and is of the same order as that of the corresponding electron bands. From dimensional analysis, one can estimate an upper bound for the orbital magnetization in flat bands, $m_z \lesssim \frac{e}{\hbar} \Delta a_0^2$, with $a_0$ the effective lattice constant in the system. The effective Land\'e $g$-factor reads $g \equiv m_z/\mu_\mathrm{B} \lesssim  \frac{2 m_e \Delta a^2}{\hbar^2}$, where $\mu_\mathrm{B} \equiv \frac{e \hbar}{2 m_e}$ is the Bohr magneton. For TBG at the first magic angle, $\Delta \sim 10 \, \mathrm{meV}$ and $a_0 \sim 200 \, \text{\r{A}}$, we estimate that the Land\'e $g$-factor can be as large as $g \sim 10^2$.

\begin{figure}[h]
\includegraphics[trim=0cm 0cm 0cm 0cm, clip=true, width=1.0\textwidth]{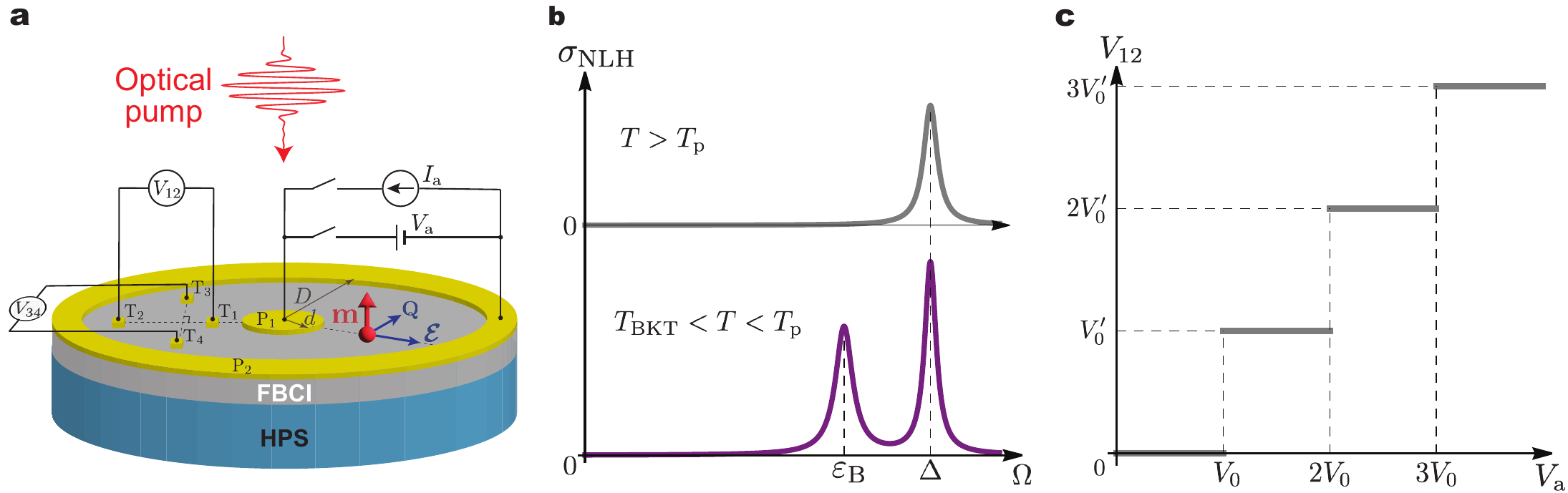}
\caption{
\textbf{Signatures of topological excitons in flat-band Chern insulators.}
\textbf{(a)} Geometry of the Corbino device for observing topological and quantum geometric properties of excitons. The Flat-band Chern insulator (FBCI) layer is fabricated on top of a high permeability substrate (HPS), which locally reduces speed of light. Electric terminals consist of a concentric hollow cylindrical metallic conductor (outer plate $\mathrm{P}_2$) encircling a solid cylindrical conductor (inner plate $\mathrm{P}_1$) attached on top of the FBCI layer. The radius of $\mathrm{P}_1$ and the inner radius of $\mathrm{P}_2$ are $d$ and $D$, respectively. Excitons can be created in the Corbino disk between $\mathrm{P}_1$ and $\mathrm{P}_2$ as linearly polarized light  is applied from the top of the device. Two terminals $\mathrm{T}_{1,2}$ ($\mathrm{T}_{3,4}$) are attached on the Corbino disk along the radial (angular) direction. The distance from $\mathrm{T}_{1(2)}$ to the disk center is $r_{1(2)}$. 
\textbf{(b)} Nonlinear Hall (NLH) conductivity $\s_\mathrm{NLH}$ as a function of light frequency $\W$ as temperature $T$ sweeps cross $T_\mathrm{p}$ [Eq.~\eqref{NLHE}]. Application of a driving current $I_\mathrm{a}$ between $\mathrm{P}_1$ and $\mathrm{P}_2$ results in a voltage drop $V_{34}$ between $\mathrm{T}_{3}$ and $\mathrm{T}_4$, which corresponds to $\s_\mathrm{NLH} = -I_\mathrm{a}/V_{34}$ up to a geometric prefactor.  
\textbf{(c)} Voltage drop quantization in exciton superfluid phase $T < T_\mathrm{BKT}$ due to the Aharonov-Casher effect. The voltage drop $V_{12}$ between $\mathrm{T}_1$ and $\mathrm{T}_2$ has the form of steps as a function of the applied voltage $V_\mathrm{a}$ between $\mathrm{P}_1$ and $\mathrm{P}_2$. The step width and height  $V_0$ 
and $V_0'$, respectively, are described in the text.} 
\label{fig-3}
\end{figure}

As in the flux quantization in a superconductor, one could investigate the topological exciton condensate between two plates of a cylindrical capacitor [see Fig.~\ref{fig-3}(a)]. The applied  electric field reads $\boldsymbol{\mathcal{E}}(\bfr) = \frac{\l}{2\pi\ep} \frac{\hat{\mathbf{e}}_r}{r}$, where $\l$ is the linear charge density in the cylinder and $\epsilon$ the dielectric constant in the medium of the condensate, resulting in a voltage drop between the two plates $\delta V \equiv V_1-V_2 = \frac{\l}{2\pi \ep} \ln (D/d)$.  For a magnetic moment $\mathbf{m} = g \mu_\mathrm{B} \hat{\mathbf{z}}$ moving on a circle of radius $r$, the Aharonov-Casher phase for a circulation reads $\theta_\mathrm{AC} = \frac{\mu \ep}{\hbar} \oint \! d \mathbf{r} \cdot \left(\mathbf{m} \times \boldsymbol{\mathcal{E}} \right) = \frac{g \mu_\mathrm{B} \l \mu}{\hbar} = 2\pi N$. Therefore, the electric field is screened inside the superfluid, with  a quantized effective linear charge density $\l = N \frac{2\pi \hbar}{g \mu_\mathrm{B} \mu}$. The voltage drop between $\mathrm{T}_1$ and $\mathrm{T}_2$ [see Fig.~\ref{fig-3}(a)] changes in integer steps $\Delta V = N V_0'$, with $V_0'= \frac{\hbar c^2}{g \mu_\mathrm{B}} \ln (r_2/r_1) \sim \frac{\td{c}^2}{g}\ln (r_2/r_1) \times 10^6 \, \mathrm{V}$, where $r_{1(2)}$ is the distance from $\mathrm{T}_{1(2)}$ to the disk center.  $c \equiv 1/\sqrt{\ep \mu}$ is the speed of light in the exciton condensate, and $\td{c} = c/c_0$ is the reduced speed of light, with $c_0$ the speed of light in vacuum. In order to observe quantization steps, we require significantly small values of $\frac{\td{c}^2}{g}\ln (r_2/r_1)$. For practical purposes, $\ln(r_2/r_1) \sim 1$, $\td{c} \sim 10^{-2}$ on  substrates with large magnetic permeability $\mu \sim 10^4$ such as iron, and $g \sim 10^2$ in 2D systems with large unit cell, so that $V_0 \sim 1 \, \mathrm{V}$. We expect that the voltage quantization shown in Fig.~\ref{fig-3}(c) is observable in the exciton superfluid phase. 

\section{Discussion}
Moire heterostructures of graphene and TMDs offer a possible platform for the observation of topological excitons ~\cite{Sharpe, Serlin}.
The most common experimentally studied samples with excitonic effects combine a semiconductor with a substrate. In the effective medium approximation, the dielectric constant of the semiconductor $\ep$ largely exceeds the dielectric constant of the substrate and produces mostly dielectric screening. As a result, the effective Coulomb interaction takes the form of $1/r$, where $r$ is the e-h relative distance, when $r$ is much larger than the dielectric screening length $r_0$. For $r \ll r_0$, the Coulomb interaction is proportional to $\ln{r}$ \cite{Rytova1967,Keldysh1979}.  The Wannier equation \eqref{wannier} in the presence of Coulomb interactions is expected to result in a series of midgap exciton states with energy spectrum that strongly deviates from that of the hydrogen atom, since the effective mass approximation is not applicable in flat bands.   We note that the specific form of the Coulomb interaction will change the energy spectrum of the excitons, but will have no effect in the topology  of the excitonic ground state. 


The electric dipole moment of  excitons $\boldsymbol{\ell}_\nu$ [Eq.~\eqref{ex-dipole}] can be very different compared to the electric dipole moment $\mathbf{P}_{\mathrm{cv}}$  of the e-h excitations at any given momentum. The optical selection rules for exciton states in flat Chern  bands can strongly depend on the vorticity of the exciton profile function $\mathcal{R}_{\nu,\mathbf{q}}$. This is distinct from the exciton selection rules in conventional materials, where the effective mass approximation is valid. In that approximation, $\boldsymbol{\ell}_\nu \approx |\wtd{\mathcal{R}}_{\nu,0}|^2 \mathbf{P}_{\mathrm{cv}}$, where $\wtd{\mathcal{R}}_{\nu,0} = \sum_{\bk} \mathcal{R}_{\nu,\bk}$  is a constant
and $\mathbf{P}_{\mathrm{cv}}$ takes the value at the band extremes~\cite{Haug1993}. For example, in TMD materials, circularly polarized light selects  e-h excitations as well as excitons in only one of the two valleys~\cite{Wang2018}. In general, quantum geometry and topological constraints can dictate distinct selection rules for excitons and e-h excitations. 

We finally remark that exciton superfluidity can also be detected via photoluminescence. At the onset of superfluidity, the spectrally integrated photoluminescence intensity can be substantially enhanced, while the photon statistics revealed by the emission can strongly deviate from a Poissonian distribution, revealing a bunching  transition~\cite{Wang2019}. We expect that radiative decay of topological excitons should lead to the spontaneous emission of circularly polarized light at the exciton resonance frequency. In order to observe the exciton superfluidity, the superfluid coherence time, limited by the light pump, should be larger than the radiative life time, which again depends on the exciton dipole moment and energy spectrum. We leave the exploration of these directions to a future study~\cite{XieUchoa2023}.

\section{Methods}

\subsection{Keldysh theory for light-pumped excitons}
Diagonalization of the single-particle Hamiltonian by a unitary matrix $\hat{U}_\bk$ results in the Hamiltonian
\begin{align} \label{ham-s}
\mathcal{H}_S(t) = & \sum_{\bk} \psi_\bk^\dagger \hat{\mathcal{H}}_\bk(t) \psi_\bk  - \frac{1}{2 A} \sum_{a_{1},\ldots, a_4} \sum_{\bk_1,\ldots \bk_4}  w_{\bk_1,\bk_2,\bk_3,\bk_3+\bk_2 -\bk_1}^{a_1,a_2,a_3,a_4} \nn \\
& \times \psi_{a_1,\bk_1}^\dagger\psi_{a_4,\bk_3+\bk_2-\bk_1}^\dagger \psi_{a_2,\bk_2} \psi_{a_3,\bk_3}.
\end{align}     
where $a_{1,\cdots, 4} = \mathrm{c,v}$. The effective Coulomb interaction reads    
\be \label{V0}
w_{\bk_1,\bk_2,\bk_3\bk_4}^{a_1,a_2,a_3,a_4} = v_{\bk_3-\bk_1} \mathcal{U}^{a_1,a_3}_{\bk_1,\bk_3} \mathcal{U}^{a_4,a_2}_{\bk_4,\bk_2},
\ee
In low-temperature insulators, the effective interaction \eqref{V0} is dominated by the exciton channel $w_{\bk_1,\bk_2,\bk_3,\bk_4}^{a,\bar{a},a,\bar{a}}$ and $w_{\bk_1,\bk_2,\bk_3,\bk_4}^{a,a,\bar{a},\bar{a}}$ where $\bar{a} = \mathrm{v}$ ($\mathrm{c}$) for $a = \mathrm{c}$ ($\mathrm{v}$). Therefore, the interaction Hamiltonian in Eq.~\eqref{ham-s} can be simplified into Eq.~\eqref{ham-s-eff}.  
For simplicity, we assume that the reservoir is composed of noninteracting electrons in equilibrium with a constant density of states, and that each site of the system uniformly couples to the reservoir via tunneling processes. The reservoir electrons obeys the Fermi-Dirac distribution function $f_F(\ep) = 1/[e^{(\ep-\mu)/T}+1]$. 

We write the Keldysh action of the light-driven interacting two-band system as $\mathcal{S} = \mathcal{S}_0 + \mathcal{S}_I$, where 
\begin{align}
\mathcal{S}_0 = \,& \iint dt_1 dt_2 \sum_{\bk}  \bar{\psi}_{\bk}(t_1) \, \breve{G}_\bk^{-1}(t_1,t_2) \, \psi_{\bk}(t_2), \label{S0-0} \\
\mathcal{S}_I = \,& \sum_{\k = \pm 1} \k  \int dt \sum_{\bk_1, \dots, \bk_4} \mathsf{W}_{\bk_1,\bk_2;\bk_3,\bk_4} \bar{\Phi}_{\k,\bk_1,\bk_2}(t) \Phi_{\k,\bk_3,\bk_4}(t), \label{S1-0}
\end{align}
where $\breve{G}_\bk^{-1}(t_1,t_2)$ is the single-particle Green's function incorporating the light field and the fermionic bath. The dipole-dipole interaction potential is defined as $\mathsf{W}_{\bk_1,\bk_2;\bk_3,\bk_4} = \frac{1}{A}\del_{\bk_1-\bk_2,\bk_3-\bk_4} W_{\frac{\bk_1+\bk_2}{2},\frac{\bk_3+\bk_4}{2};\bk_1-\bk_2}$ via Eq.~\eqref{Weff}.  

We introduce the complex Hubbard-Stratonovich fields  $X_{\k,\bk,\bp}(t)$ and $X^\ast_{\k,\bk,\bp}(t)$ as order parameters to decouple the interaction in Eq.~\eqref{S1-0}
~\cite{Kamenev2011}. In time-contour space, we apply the Keldysh-Larkin-Ovchinikov rotations $R$ and $\bar{R}$ to the fermionic fields $\{\psi,\bar{\psi}\}$ and bosonic fields $\{\bar{X},X\}$, i.e., $\psi(t) \to R \psi(t)$ and $\bar{\psi}(t) = \bar{\psi}(t) \bar{R}$, and $X(t) \to \sqrt{2} R X(t)$ and $\bar{X}(t) \to \sqrt{2} \bar{X}(t) R$, where $R = (\k_1 + \k_3)/\sqrt{2}$ and $\bar{R} =(1 - i \k_2)/\sqrt{2}$, with $\k_{1,2,3}$ being the Pauli matrices in time-contour space. Integrating out the fermionic fields, we obtain the effective action for the bosonic fields 
\begin{align} \label{S_eff}
\mathcal{S}_\mathrm{eff}[X^\ast,X; V^\ast,V] = & -\int\!dt \, \bar{X}(t) \left( \k_1 \otimes \hat{\mathsf{W}}^{-1} \right) X(t) \nn \\
                               & -i \mathrm{Tr} \ln \left[1- \breve{G} (\breve{X}  + \breve{V}) \right],
\end{align}
where for a matrix field $\breve{M}$ with elements $M_{\k\k',aa',\bk\bk'}(t,t')$ we have defined $\mathrm{Tr} \breve{M} = \sum_{\k=1,2}\sum_{a=\mathrm{c,v}}\sum_{\bk}\int\!dt M_{\k\k,aa,\bk\bk}(t,t)$, and $\breve{X}_{\bk_1,\bk_2}(t_1,t_2) \equiv \delta(t_1-t_2) \left( \k_0 \otimes \hat{X}_{\bk_1,\bk_2}^{(1)}(t_2) +  \k_1 \otimes \hat{X}_{\bk_1,\bk_2}^{(2)}(t_2)\right)$, with $\hat{X}_{\bk_1,\bk_2}^{(\k)}(t) = X_{\k,\bk_1,\bk_2}(t) \s_{+}  + X_{\k,\bk_1,\bk_2}^\ast (t) \s_{-}$.  $\s_{\pm} \equiv (\s_{1} \pm i \s_2)/2$ are the Pauli matrices in band space. $\breve{V}$ takes the same matrix structure as that of $\breve{X}$ incorporating the source fields $V_{\k,\bk,\bp}(t)$ and $V^\ast_{\k,\bk,\bp}(t)$ that generate dipole-dipole correlation functions.

In Eq.~\eqref{S_eff}, the single-particle Green's function satisfies the periodicity $\breve{\mathcal{G}}_\bk (t_1+T_0,t_2+T_0) = \breve{\mathcal{G}}_\bk(t_1,t_2)$ where $T_0 = 2 \pi/\W$ is the period of the light field. Its  Fourier transform is $\breve{\mathcal{G}}_{\bk}(t_1,t_2) = \sum_{m,n = \mathbb{Z}} \int_{\w} e^{-i(\w + m \W)t_1} e^{i(\w + n \W)t_2} \breve{\mathcal{G}}_{\bk,mn}(\w)$, where we have defined $\int_\w \equiv \int_{-\W/2}^{\W/2} \! \frac{d\w}{2\pi}$ as the integral over first Floquet zone. We have
\be 
\breve{\mathcal{G}}_\bk(\w) = \begin{pmatrix} \hat{R}_\bk(\w) & \hat{K}_\bk(\w) \\ 0 & \hat{A}_\bk(\w)  \end{pmatrix}^{-1},
\ee 
which contains the Keldysh, Floquet, and band spaces. The retarded component reads $\hat{R}_\bk^{-1}(\w) = \hbar(\w + i \G/2) - \hat{\mathcal{H}}_{F,\bk}$, where $\hat{\mathcal{H}}_{F,\bk,mn} \equiv \hat{\mathcal{H}}_{\bk,m-n} - n \hbar \W \delta_{mn}$ is the Floquet Hamiltonian, with $\hat{\mathcal{H}}_{\bk,m} = \frac{1}{T_0} \int_{0}^{T_0} dt \hat{\mathcal{H}}_\bk(t) e^{i m \W t}$, and $\G = 2\pi |\g|^2 \nu/\hbar$  the effective single-particle decay rate,  $\g$ the tunnelling strength between the system and the bath and $\nu$ the bath DOS. The advanced component satisfies $\hat{A}_\bk(\w) = \hat{R}_\bk^\dagger(\w)$. The Keldysh component reads $\hat{K}_\bk(\w) = \hat{R}_\bk(\w) \hat{\Sigma}(\w) \hat{A}_\bk(\w)$, where $\hat{\Sigma}(\w) = -i \hbar \G (1 - 2 \hat{f}_F(\w))$ is the Keldysh component of the self-energy arising from the coupling to the bath.  The matrix $f_{F,mn}(\w) = \delta_{mn} f_{F}(\w+m\W)$ is diagonal in Floquet space and encodes the Fermi-Dirac distribution function of the bath. Moreover, the observables are directly related to the lesser Green's function $\hat{G}_\bk^<(\w) = ( \hat{K}_\bk(\w) - \hat{R}_\bk(\w) + \hat{A}_\bk(\w) )/2$.

In the absence of the source fields $V=V^\ast=0$, Eq.~\eqref{S_eff} gives the effective action of the order parameters. Taking the saddle point $\del S_\mathrm{eff} [X,\bar{X}]/\del X_{\bk_1,\bk_2}^{(2),\ast}(t) |_{X^{(2)}=0} =0$, we obtain the classical equations of motion (EOM) for the order parameter $X_{\bk_1,\bk_2} (t) \equiv X_{\k=1,\bk_1,\bk_2} (t)$,
\be \label{gap-eq}
\sum_{\bk',\bk''} \hat{\mathsf{W}}^{-1}_{\bk_1,\bk_2;\bk',\bk''} X_{\bk',\bk''} (t) = i \mathcal{K}_{\mathrm{cv},\bk',\bk''}(t,t;[X]),
\ee
and the complex conjugation, where $\hat{\mathcal{K}}$ is the Keldysh component of the Green's function $\breve{\mathcal{G}}$ of the saddle-point Hamiltonian 
\be \label{mf-ham-0}
\underline{\mathcal{H}}_{\bk_1,\bk_2}(t) = \mathcal{H}_{\bk_1}(t) \del_{\bk_1,\bk_2} + X_{\bk_1,\bk_2}(t) \s_{+} + X_{\bk_1,\bk_2}^\ast(t) \s_{-}.
\ee
We obtain $\breve{\mathcal{G}} = ( \breve{G}^{-1} - \breve{X} )^{-1} \equiv \begin{pmatrix} \hat{\mathcal{R}} & \hat{\mathcal{K}} \\ 0 & \hat{\mathcal{A}}  \end{pmatrix}$, where $\breve{X}_{\bk_1,\bk_2}(t_1,t_2) = \delta(t_1-t_2) \k_0 \otimes ( X_{\bk_1,\bk_2}(t) \s_{+}  + X_{\bk_1,\bk_2}^\ast (t) \s_{-} )$. Equations \eqref{gap-eq} and \eqref{mf-ham-0} form the self-consistent dynamical equations for excitons. 

Expanding the EOM \eqref{gap-eq} up to linear order in the order parameter, we obtain the linear gap equation for the excitons,
\be \label{sol-gap}
X_{\bk_1,\bk_2} (t) = i\sum_{\bk'} \int dt' \left( \hat{\mathsf{W}}^{-1} - \hat{\pi}^R \right)^{-1}_{\bk_1,\bk_2;\bk',\bk'}(t,t') K_{\mathrm{cv},\bk'}(t',t'),
\ee
where 
\begin{align}
& \pi^R_{\bk_1,\bk_2,\bk_3,\bk_4}(t_1,t_2) = \frac{i}{2} \del_{\bk_1,\bk_3} \del_{\bk_2,\bk_4} \nn \\
& \times \mathrm{Tr}_\s \big[ \s_- \hat{R}_{\bk_1}(t_1,t_2) \s_+ \hat{K}_{\bk_2}(t_2,t_1)  + \s_- \hat{K}_{\bk_1}(t_1,t_2) \s_+ \hat{A}_{\bk_2}(t_2,t_1)  \big]
\end{align}
 is the retarded interband polarization function. 
The Keldysh component of the interband Green's function $K_{\mathrm{cv},\bk}$ is the source that generates the order parameter. Conversely,  $K_\mathrm{cv,\bk} =0$ in the absence of the light pump.  

The generating function of the interband electric dipole correlation functions is defined by $Z[V^\ast,V] \equiv \int \! \mathcal{D}\bar{X} \mathcal{D}X \, e^{i S_\mathrm{eff}[X^\ast,X; V^\ast,V]}$. In particular, we calculate the dipole propagator $\Pi^{\k_1 \k_2}(1,2) \equiv  -\frac{i}{2} \left. \frac{ \del^2 \ln Z}{ \del V_{\k_1}^{\ast}(1) \del V_{\k_2}(2)} \right|_{V=0}$ and obtain
\begin{align}
\Pi^{\k_1 \k_2}(1,2) = & \left\langle \pi^{\k_1\k_2}(1,2) \right \rangle \nn \\
                                  & + \frac{i}{2} \left( \left\langle p_-^{\k_1}(1) p_+^{\k_2}(2) \right\rangle - \left\langle p_-^{\k_1}(1)\right \rangle \left\langle p_+^{\k_2}(2) \right \rangle \right),
\end{align} 
where ``$1$'' and ``$2$'' are short-hand notations for $(\bk_1,\bk_2,t_1)$ and $(\bk_3,\bk_4,t_2)$, respectively, $\left\langle \cdots \right\rangle \equiv \int \! \mathcal{D} X^\ast \mathcal{D} X \cdots  e^{i \mathcal{S}_\mathrm{eff}[X^\ast,X]}$, and
\begin{align}
\pi^{\k_1\k_2}(1,2) =  &\, \frac{i}{2} \mathrm{Tr}_{\k,\s} \big[ \breve{\mathcal{G}}_{\bk_4,\bk_2}(t_2,t_1;[X^\ast,X]) ( \g_{\k_1} \otimes \s_- ) \nn \\
                                  & \times \breve{\mathcal{G}}_{\bk_1,\bk_3}(t_1,t_2;[X^\ast,X]) ( \g_{\k_2} \otimes \s_+ ) \big], \label{G-G} \\
        p_-^{\k_1}(1)  =  &\, i \mathrm{Tr}_{\k,\s} \left[ \breve{\mathcal{G}}_{\bk_1,\bk_2}(t_1,t_1;[X^\ast,X])  \left( \g_{\k_1} \otimes \s_- \right) \right], \\ 
        p_+^{\k_2}(2) =  &\, i \mathrm{Tr}_{\k,\s} \left[ \breve{\mathcal{G}}_{\bk_4,\bk_3}(t_2,t_2;[X^\ast,X]) \left( \g_{\k_2} \otimes \s_+ \right) \right], 
\end{align}    
with $\g_{1,2} = \k_{0,1}$, respectively.

Up to leading order effects of the light pump, the order parameter $X$ is proportional to the light field. On the right-hand side of Eq.~\eqref{sol-gap}, it is sufficient to take the pump-free retarded polarization function. In frequency space, it takes the expression 
$\pi^R_{\bk_1,\bk_2,\bk_3,\bk_4}(\w) = -\del_{\bk_1,\bk_3} \del_{\bk_2,\bk_4} \del{f}_{\bk_1\bk_2}/(\w - \Delta_{\bk_1,\bk_2} + i\G)$. The source term is proportional to the light field $K_{\mathrm{cv},\bk}(t,t) =  i e^{-i\W t} \del{f}_{\bk_1\bk_2} (\mathbf{E} \cdot \mathbf{P}_{\mathrm{cv},\bk})/(\W - \Delta_{\bk_1\bk_2}  + i \G)$. Finally, we obtain Eq.~\eqref{X-q} from Eq.~\eqref{sol-gap}. See Supplementary Information for details. 

\subsection{Flat-band Haldane model}

In general a two-band flat-band model reads $H_{\bk} = \frac{\Delta}{2} \hat{\mathbf{d}}(\bk) \cdot \boldsymbol{\s}$ where $\hat{\mathbf{d}}(\bk)$ is a unit vector, i.e., $|\hat{\mathbf{d}}(\bk)|=1$. Exploiting the Haldane model on honeycomb lattice, we introduce the $\mathbf{d}$-vector
\begin{align} \label{2b-haldane}
& d_1(\bk) = t_1 \sum_j \cos(\bk \cdot \mathbf{a}_j), \quad 
d_2(\bk) = t_1 \sum_j \sin(\bk\cdot \mathbf{a}_j), \nn \\
& d_3(\bk) = -t_2 \sum_j \sin(\bk\cdot \mathbf{b}_j),
\end{align}    
where $t_{1,2}$ are real parameters and $\mathbf{a}_1 = a(1,0)$, $\mathbf{a}_2=a(-1/2,\sqrt{3}/2)$, $\mathbf{a}_3 = - \mathbf{a}_1 - \mathbf{a}_2$, $\mathbf{b}_1=\mathbf{a}_2 - \mathbf{a}_3$, $\mathbf{b}_2=\mathbf{a}_3 - \mathbf{a}_1$, $\mathbf{b}_3=\mathbf{a}_1 - \mathbf{a}_2$. We define $\hat{\mathbf{d}}(\bk) = \mathbf{d}(\bk)/|\mathbf{d}(\bk)|$. The conduction and valence bands take the energies $\ep_a = \s_a \Delta/2$ and the eigenstates $\psi_{a}(\bk) = \frac{1}{\sqrt{2 ( 1+\s_a \hat{d}_3(\bk) )}} \begin{pmatrix} \s + \hat{d}_3(\bk) &  \hat{d}_1(\bk) + i \hat{d}_2(\bk) \end{pmatrix}^T$, with $\s_{\mathrm{c(v)}}=+1\, (-1)$. For this mode $\mathcal{C}_\mathrm{c}=-\mathcal{C}_\mathrm{v}=1$. Numerically solving the Wannier equation \eqref{wannier} for a constant interaction, we obtain a single exciton bound state and the wave function is shown in Fig.~\ref{fig-2}(c) for $t_1=t_2=1$.

\section{Acknowledgements}
We thank K. Mullen, M. Foster, A. Passupathy, V. Menon, and M. Horvath for stimulating discussions. The work of H.-Y.X. is supported by the Dodge Family Fellowship granted by the University of Oklahoma. P.G. acknowledges support from NSF-DMR-2130544, NSF-DMR2037996 and infrastructural support from NSF HRD-2112550 (NSF CREST Center IDEALS). M.M. was supported by the U.S. Department of Energy, Office of Basic Energy Sciences, Early Career Award Program, under Award No. DE-SC0022883. B.U. acknowledges NSF grant DMR-2024864 for support.

\newpage

\newpage

\newpage

\clearpage



\section*{Supplementary material}

# Theory of topological exciton insulators and condensates in flat Chern bands

Hong-Yi Xie,[1] Pouyan Ghaemi,[2] Matteo Mitrano,[3] and Bruno Uchoa[1]

[1] *Department of Physics and Astronomy, Center for Quantum Research and Technology,
University of Oklahoma, Norman, Oklahoma 73069, USA*

[2] *Department of Physics, City University of New York, New York, 73069, USA*

[3] *Department of Physics, Harvard University, Cambridge, Massachussets, USA*

**SINGLE-PARTICLE GREEN'S FUNCTION**

According to Eq. (3) in the main text, up to linear order of the light field, the Hamiltonian reads

$$\hat{\mathcal{H}}_{\mathbf{k}}(t) = \hat{\varepsilon}_{\mathbf{k}} - \left( \mathbf{E} e^{-i\Omega t} - \mathbf{E}^* e^{i\Omega t} \right) \cdot \hat{\mathbf{P}}_{\mathbf{k}}, \quad \hat{\mathbf{P}}_{\mathbf{k}} = \frac{e}{\Omega} \begin{pmatrix} i\partial_{\mathbf{k}}\varepsilon_{c,\mathbf{k}} & \Delta_{\mathbf{k}} \mathcal{A}_{cv,\mathbf{k}} \\ -\Delta_{\mathbf{k}} \mathcal{A}_{cv,\mathbf{k}}^* & i\partial_{\mathbf{k}}\varepsilon_{v,\mathbf{k}} \end{pmatrix}. \tag{S1}$$

In Floquet space the Hamiltonian S1 takes the form $\hat{\mathcal{H}}_{F,\mathbf{k},mn} = \delta_{mn} \left( \hat{\varepsilon}(\mathbf{k}) - n\Omega \right) - \left( \delta_{m,n+1}\mathbf{E} - \delta_{m,n-1}\mathbf{E}^* \right) \cdot \hat{\mathbf{P}}_{\mathbf{k}}$. About the optical resonance $\Omega \sim \Delta_{\mathbf{k}}$, the leading order degenerate perturbation in $\mathbf{E}$ and $\mathbf{E}^*$ solves the block-diagonal Hamiltonian

$$\hat{\mathcal{H}}_{F,\mathbf{k},mn} = \delta_{mn} \left( \hat{\varepsilon}_{\mathbf{k}} - n\Omega \right) + \delta_{m,n+1}\mathbf{E} \cdot \mathbf{P}_{cv,\mathbf{k}}\hat{\sigma}_+ + \delta_{m,n-1}\mathbf{E}^* \cdot \mathbf{P}_{cv,\mathbf{k}}^*\hat{\sigma}_-, \tag{S2}$$

which gives Eq. (4) in time domain. The eigenstates of the Hamiltonian S2 read $\hat{\mathcal{H}}_{F,\mathbf{k}}\hat{O}_{\mathbf{k}} = \hat{O}_{\mathbf{k}}\hat{\mathcal{E}}_{\mathbf{k}}$, where

$$\hat{\mathcal{E}}_{\mathbf{k},mn} = \delta_{mn}\text{diag}\{\mathcal{E}_{c,\mathbf{k},m}, \mathcal{E}_{v,\mathbf{k},m}\}, \quad \mathcal{E}_{c(v),\mathbf{k},m} = \left( \frac{\varepsilon_{c,\mathbf{k}} + \varepsilon_{v,\mathbf{k}}}{2} - m\Omega \right) - (+) \left( |\mathbf{d}_{\mathbf{k}}| - \frac{\Omega}{2} \right), \tag{S3}$$

$$\hat{O}_{\mathbf{k},mn} = \delta_{mn}\cos(\phi_{\mathbf{k}}/2)e^{i\hat{\sigma}_3\varphi_{\mathbf{k}}/2} + \delta_{m,n+1}\sin(\phi_{\mathbf{k}}/2)e^{i\varphi_{\mathbf{k}}/2}\hat{\sigma}_+ - \delta_{m,n-1}\sin(\phi_{\mathbf{k}}/2)e^{-i\varphi_{\mathbf{k}}/2}\hat{\sigma}_-, \tag{S4}$$

with $\mathbf{d}_{\mathbf{k}} = (d_{1,\mathbf{k}}, d_{2,\mathbf{k}}, d_{3,\mathbf{k}})$ and $d_{1,\mathbf{k}} = \text{Re}(\mathbf{E} \cdot \mathbf{P}_{cv,\mathbf{k}})$, $d_{2,\mathbf{k}} = \text{Im}(\mathbf{E} \cdot \mathbf{P}_{cv,\mathbf{k}})$, $d_{3,\mathbf{k}} = (\varepsilon_{v,\mathbf{k}} - \varepsilon_{c,\mathbf{k}} + \Omega)/2$, and the SU(2) angles $\varphi_{\mathbf{k}} = \text{Arg}(\mathbf{E} \cdot \mathbf{P}_{cv,\mathbf{k}})$ and $\phi_{\mathbf{k}} = \arccos(d_{3,\mathbf{k}}/|\mathbf{d}_{\mathbf{k}}|)$. At resonance $\Omega = \Delta_{\mathbf{k}}$, $d_{3,\mathbf{k}} = 0$ and the Rabi splitting $\Omega_{R,\mathbf{k}} = 2e|\mathbf{E} \cdot \mathcal{A}_{cv,\mathbf{k}}|$.

The single-particle Green's functions are block-diagonal in Floquet space and the nonvanishing blocks read

$$\begin{pmatrix} R_{\mathbf{k},m,v;m,v}(\omega) & R_{\mathbf{k},m,v;m+1,c}(\omega) \\ R_{\mathbf{k},m+1,c;m,v}(\omega) & R_{\mathbf{k},m+1,c;m+1,c}(\omega) \end{pmatrix} = \hat{O}_{\mathbf{k}} \begin{pmatrix} \frac{1}{\omega - \mathcal{E}_{m,v,\mathbf{k}} + i\Gamma/2} & 0 \\ 0 & \frac{1}{\omega - \mathcal{E}_{m+1,c,\mathbf{k}} + i\Gamma/2} \end{pmatrix} \hat{O}_{\mathbf{k}}^\dagger, \quad \hat{A}_{\mathbf{k}}(\omega) = \hat{R}_{\mathbf{k}}^\dagger(\omega), \tag{S5}$$

$$\begin{pmatrix} K_{\mathbf{k},m,v;m,v}(\omega) & K_{\mathbf{k},m,v;m+1,c}(\omega) \\ K_{\mathbf{k},m+1,c;m,v}(\omega) & K_{\mathbf{k},m+1,c;m+1,c}(\omega) \end{pmatrix}$$
$$= \hat{O}_{\mathbf{k}} \begin{pmatrix} -i\pi \left[ f_{m,+}(\omega) - f_{m,-}(\omega)\cos\phi_{\mathbf{k}} \right] \mathcal{L}\left( \omega; \mathcal{E}_{m,v,\mathbf{k}}, \frac{\Gamma}{2} \right) & -\frac{i(\hbar\Gamma/2)f_{m,-}(\omega)\sin\phi_{\mathbf{k}}}{[\omega - \mathcal{E}_{m,v,\mathbf{k}} + i\Gamma/2][\omega - \mathcal{E}_{m+1,c,\mathbf{k}} - i\Gamma/2]} \\ -\frac{i(\hbar\Gamma/2)f_{m,-}(\omega)\sin\phi_{\mathbf{k}}}{[\omega - \mathcal{E}_{m,v,\mathbf{k}} - i\Gamma/2][\omega - \mathcal{E}_{m+1,c,\mathbf{k}} + i\Gamma/2]} & -i\pi \left[ f_{m,+}(\omega) + f_{m,-}(\omega)\cos\phi_{\mathbf{k}} \right] \mathcal{L}\left( \omega; \mathcal{E}_{m+1,c,\mathbf{k}}, \frac{\Gamma}{2} \right) \end{pmatrix} \hat{O}_{\mathbf{k}}^\dagger, \tag{S6}$$

where the SU(2) matrix reads

$$\hat{O}_{\mathbf{k}} = \begin{pmatrix} \cos(\phi_{\mathbf{k}}/2)e^{-i\varphi_{\mathbf{k}}/2} & -\sin(\phi_{\mathbf{k}}/2)e^{-i\varphi_{\mathbf{k}}/2} \\ \sin(\phi_{\mathbf{k}}/2)e^{i\varphi_{\mathbf{k}}/2} & \cos(\phi_{\mathbf{k}}/2)e^{i\varphi_{\mathbf{k}}/2} \end{pmatrix}, \tag{S7}$$

which encodes the eigenstates of the Floquet Hamiltonian, the effective distribution functions reads $f_{m,+}(\omega) = 2\left[ 1 - f_F(\omega + m\Omega) - f_F(\omega + (m+1)\Omega) \right]$ and $f_{m,-}(\omega) = 2\left[ f_F(\omega + m\Omega) - f_F(\omega + (m+1)\Omega) \right]$, and $\mathcal{L}(\omega; \omega_0, \gamma) = \frac{1}{\pi} \frac{\gamma}{(\omega - \omega_0) + \gamma^2}$

is the Lorentzian. As a result, the lesser Green's function reads

$$
\begin{pmatrix} G^<_{\mathbf{k},m,\mathrm{v};m,\mathrm{v}}(\omega) & G^<_{\mathbf{k},m,\mathrm{v};m+1,\mathrm{c}}(\omega) \\ G^<_{\mathbf{k},m+1,\mathrm{c};m,\mathrm{v}}(\omega) & G^<_{\mathbf{k},m+1,\mathrm{c};m+1,\mathrm{c}}(\omega) \end{pmatrix}
$$
$$
= \hat{\mathrm{O}}_{\mathbf{k}} \begin{pmatrix} i\pi \left[ g_{m,+}(\omega) - g_{m,-}(\omega) \cos\phi_{\mathbf{k}} \right] \mathcal{L}\left(\omega;\mathcal{E}_{m,\mathrm{v},\mathbf{k}},\frac{\Gamma}{2}\right) & \frac{i(\hbar\Gamma/2)g_{m,-}(\omega)\sin\phi_{\mathbf{k}}}{[\omega-\mathcal{E}_{m,\mathrm{v},\mathbf{k}}+i\Gamma/2][\omega-\mathcal{E}_{m+1,\mathrm{c},\mathbf{k}}-i\Gamma/2]} \\ \frac{i(\hbar\Gamma/2)g_{m,-}(\omega)\sin\phi_{\mathbf{k}}}{[\omega-\mathcal{E}_{m,\mathrm{v},\mathbf{k}}-i\Gamma/2][\omega-\mathcal{E}_{m+1,\mathrm{c},\mathbf{k}}+i\Gamma/2]} & i\pi \left[ g_{m,+}(\omega) + g_{m,-}(\omega) \cos\phi_{\mathbf{k}} \right] \mathcal{L}\left(\omega;\mathcal{E}_{m+1,\mathrm{c},\mathbf{k}},\frac{\Gamma}{2}\right) \end{pmatrix} \hat{\mathrm{O}}^\dagger_{\mathbf{k}}, \quad \text{(S8)}
$$

where $g_{m,\pm}(\omega) = f_F(\omega + (m+1)\Omega) \pm f_F(\omega + m\Omega)$.

In particular, the source term of the order parameter EOM 25 involves the interband component of equal-time Keldysh Green's function. From Eq. (S6) we obtain

$$
K_{\mathbf{k},\mathrm{cv}}(t,t) = -e^{-i\Omega t} \frac{i}{4} e^{i\varphi_{\mathbf{k}}} \sin\phi_{\mathbf{k}} \left\{ \left[ f_+(\mathcal{E}_{0,\mathrm{v},\mathbf{k}}) - f_+(\mathcal{E}_{1,\mathrm{c},\mathbf{k}}) \right] - \left[ f_-(\mathcal{E}_{0,\mathrm{v},\mathbf{p}}) + f_-(\mathcal{E}_{1,\mathrm{c},\mathbf{k}}) \right] \right\} \frac{|\mathbf{d}_{\mathbf{k}}|(d_{3,\mathbf{k}} - i\Gamma/2)}{|\mathbf{d}_{\mathbf{k}}|^2 + (\Gamma/2)^2} \right\}, \quad \text{(S9)}
$$

where $f_+(\omega) = 2\left[1 - f_F(\omega) - f_F(\omega + \Omega)\right]$ and $f_-(\omega) = 2\left[f_F(\omega) - f_F(\omega + \Omega)\right]$. Up to linear order in light field, we obtain the source term in Methods.

## ELECTRIC DIPOLE CORRELATION FUNCTIONS

Up to leading order in light field, we only need the pump-free dipole-dipole correlation functions. Expanding the effective action in Eq. (21) to quadratic order in order parameter and in source field, we obtain the RPA action

$$
S_{\mathrm{RPA}}[\bar{X},X;\bar{V},V] = \int \frac{d\omega}{2\pi} \bar{X}(\omega)\left(\breve{\pi}(\omega) - \hat{\kappa}_1 \otimes \hat{W}^{-1}\right)X(\omega) + \bar{X}(\omega)\breve{\pi}(\omega)V(\omega) + \bar{V}(\omega)\breve{\pi}(\omega)X(\omega) + \bar{V}(\omega)\breve{\pi}(\omega)V(\omega), \quad \text{(S10)}
$$

where $\breve{\pi} = \begin{pmatrix} 0 & \hat{\pi}^A \\ \hat{\pi}^R & \hat{\pi}^K \end{pmatrix}$ is the interband polarization function with the components

$$
\hat{\pi}^{R(A)}_{\mathbf{k}_1,\mathbf{k}_2,\mathbf{k}_3,\mathbf{k}_4}(\omega) = -\frac{\delta_{\mathbf{k}_1,\mathbf{k}_3}\delta_{\mathbf{k}_2,\mathbf{k}_4}\delta f_{\mathbf{k}_1,\mathbf{k}_2}}{\omega - \Delta_{\mathbf{k}_1,\mathbf{k}_2} \pm i\Gamma}, \quad \hat{\pi}^K_{\mathbf{k}_1,\mathbf{k}_2,\mathbf{k}_3,\mathbf{k}_4}(\omega) = \coth\left(\frac{\omega}{2T}\right)\left[\hat{\pi}^R(\omega) - \hat{\pi}^A(\omega)\right]. \quad \text{(S11)}
$$

The generating function reads $\ln Z_{\mathrm{RPA}}[\mathsf{V}^*,\mathsf{V}] = i\int \frac{d\omega}{2\pi} \bar{V}(\omega)\breve{\Pi}(\omega)V(\omega)$, where the dipole-dipole correlation function $\breve{\Pi} = \begin{pmatrix} 0 & \hat{\Pi}^A \\ \hat{\Pi}^R & \hat{\Pi}^K \end{pmatrix}$ takes the components

$$
\hat{\Pi}^{R(A)}(\omega) = \left(\hat{\pi}^{R(A),-1}(\omega) - \hat{W}\right)^{-1}, \quad \hat{\Pi}^K(\omega) = \coth\left(\frac{\omega}{2T}\right)\left(\hat{\Pi}^R(\omega) - \hat{\Pi}^A(\omega)\right). \quad \text{(S12)}
$$

The energy poles of the retarded component $\hat{\Pi}^R(\omega)$ defines the exciton Hamiltonian 6 and takes the spectrum representation in Eq. (8).

## QUADRATIC APPROXIMATION OF EXCITON BANDS

Expanding the effective interaction in Eq. (6) up to quadratic order of the COM momentum $\mathbf{Q}$, we obtain

$$
\mathcal{V}_{\alpha,\nu\nu'} = -v \sum_{\mathbf{q},\mathbf{q}'} \left( \mathcal{R}^*_{\nu,\mathbf{q}} \Xi^{\mathrm{cc},\alpha}_{\mathbf{q},\mathbf{q}'} \mathcal{R}_{\nu',\mathbf{q}'} \Xi^{\mathrm{vv}}_{\mathbf{q}',\mathbf{q}} - \mathcal{R}^*_{\nu,\mathbf{q}} \Xi^{\mathrm{cc}}_{\mathbf{q},\mathbf{q}'} \mathcal{R}_{\nu'}(\mathbf{q}') \Xi^{\mathrm{vv},\alpha}_{\mathbf{q}',\mathbf{q}} \right), \quad \text{(S13)}
$$

$$
\mathcal{M}^{-1}_{\alpha\beta,\nu\nu'} = -v \sum_{\mathbf{q},\mathbf{q}'} \left( \mathcal{R}^*_{\nu,\mathbf{q}} \Xi^{\mathrm{cc},\alpha\beta}_{\mathbf{q},\mathbf{q}'} \mathcal{R}_{\nu',\mathbf{q}'} \Xi^{\mathrm{vv}}_{\mathbf{q}',\mathbf{q}} - 2\mathcal{R}^*_{\nu,\mathbf{q}} \Xi^{\mathrm{cc},\alpha}_{\mathbf{q},\mathbf{q}'} \mathcal{R}_{\nu'}(\mathbf{q}') \Xi^{\mathrm{vv},\beta}_{\mathbf{q}',\mathbf{q}} + \mathcal{R}^*_{\nu,\mathbf{q}} \Xi^{\mathrm{cc}}_{\mathbf{q},\mathbf{q}'} \mathcal{R}_{\nu',\mathbf{q}'} \Xi^{\mathrm{vv},\alpha\beta}_{\mathbf{q}',\mathbf{q}} \right) + v\left(\frac{\Omega}{e\Delta}\right)^2 \ell_\alpha \ell^*_\beta. \quad \text{(S14)}
$$

where we have assumed constant interaction $v_{\mathbf{q}} = v$ and defined

$$\Xi_{\mathbf{q},\mathbf{q}'}^{aa,\alpha} = \frac{1}{2}\left(\partial_{q_\alpha} + \partial_{q'_\alpha}\right)\Xi_{\mathbf{q},\mathbf{q}'}^{aa}, \quad \Xi_{\mathbf{q},\mathbf{q}'}^{aa,\alpha\beta} = \frac{1}{8}\left(\partial_{q_\alpha}\partial_{q_\beta} + 2\partial_{q_\alpha}\partial_{q'_\beta} + \partial_{q'_\alpha}\partial_{q'_\beta}\right)\Xi_{\mathbf{q},\mathbf{q}'}^{aa}. \tag{S15}$$

We note that the last term in Eq. (S14) arises from the exchange interaction $W^{(e)}$ in Eq. (2).

## EFFECTIVE MAGNETIC DIPOLE MOMENTS OF EXCITONS

The effective magnetization of excitons arises from the orbital magnetization of the electron bands, the variation of Fermi-Dirac distribution, and the variation of electron wave functions. At zero temperature, we neglect the occupation variation. For the conduction and valence bands embedded in a set of flat bands, both the effective gap and the interaction term are modified by magnetic field. On one hand, the orbital magnetization of the bands reads

$$\mathbf{m}_a(\mathbf{k}) = \frac{\Delta}{\hbar}\mathsf{D}_{a,\mathbf{k}}\hat{\mathbf{z}}, \quad \mathsf{D}_{a,\mathbf{k}} = \sum_{b\neq a}\mathrm{Im}(\mathcal{A}_{ab,x,\mathbf{k}}\mathcal{A}_{ab,y,\mathbf{k}}^*), \tag{S16}$$

with $a \in \{\mathrm{c,v}\}$. The net magnetization of the e-h pair state $|e, \mathbf{k} + \mathbf{Q}\rangle|h, -\mathbf{k}\rangle$ is $\mathbf{m}_{\mathrm{c},\mathbf{k}+\mathbf{Q}} - \mathbf{m}_{\mathrm{v},\mathbf{k}}$. On the other hand, electron wave-function variation modified the effective e-h interaction via wave function overlaps. In the presence of magnetic field, the electron Hamiltonian $H = H_0 + V$, where $V = \frac{e}{2}\left[\hat{\mathbf{v}} \cdot \mathbf{A}(\mathbf{r}) + \mathbf{A}(\mathbf{r}) \cdot \hat{\mathbf{v}}\right]$, with $\hat{\mathbf{v}}$ being the electron velocity operator and $\mathbf{A}(\mathbf{r}) = \frac{1}{2}\mathbf{B} \times \mathbf{r}$ being the vector potential. First order perturbation gives the variation of the wave function $|\delta\psi_{a,\mathbf{k}}\rangle = -\frac{e}{\hbar}B\mathsf{D}_{a,\mathbf{k}}|\psi_{a,\mathbf{k}}\rangle$. According to the Wannier equation (7), for COM motion at rest, the exciton magnetic moment is given by

$$m_\nu(\mathbf{Q} = 0) = -\frac{e}{\hbar}\sum_{\mathbf{k}}|\mathcal{R}_{\nu,\mathbf{k}}|^2\sum_{a=\mathrm{c,v}}\sum_{b\neq\mathrm{c,v}}(2\varepsilon_{\mathrm{B},\nu} + \sigma_a\Delta_{ab})\mathrm{Im}(\mathcal{A}_{ab,x,\mathbf{k}}\mathcal{A}_{ab,y,\mathbf{k}}^*), \tag{S17}$$

where $\varepsilon_{\mathrm{B},\nu}$ is the binding energy, $\Delta_{ab}$ is the energy difference between bands $a$ and $b$, and $\sigma_{\mathrm{c(v)}} = +1\,(-1)$. Obviously, For a two-band model with particle-hole symmetry, the magnetic dipole moment vanishes since the orbital magnetizations of the two bands exactly compensate each other. The minimal flat-band tight-binding model that gives finite exciton magnetic moment is a three-band model.



\end{document}